\documentclass[fleqn,usenatbib,useAMS]{mnras}

\usepackage[thinc]{esdiff}
\usepackage{multirow}
\usepackage{xcolor}

\usepackage{graphicx}	
\usepackage{amsmath}	
\usepackage{amssymb}	
\usepackage{multicol}        
\usepackage{bm}		
\usepackage{pdflscape}	

\newcommand{\msun}{\(\textup{M}_\odot\)}   
\newcommand{\gcm}{g\,cm$^{-3}$}             
\newcommand*\subtxt[1]{_{\textnormal{#1}}}
\DeclareRobustCommand\_{\ifmmode\expandafter\subtxt\else\textunderscore\fi}    

\usepackage[T1]{fontenc}
\usepackage{ae,aecompl}

\usepackage{newtxtext,newtxmath}
\title[Grouped star formation]{Grouped star formation: converting sink particles to stars in hydrodynamical simulations}
\author[Liow et al.]{
Kong You Liow,$^{1}$\thanks{E-mail: kl457@exeter.ac.uk}
Steven Rieder,$^{1,2}$
Clare L. Dobbs,$^{1}$
and Sarah E. Jaffa$^{3,4}$
\\
$^{1}$School of Physics and Astronomy, University of Exeter, Stocker Road, Exeter EX4 4QL, UK\\
$^{2}$RIKEN Center for Computational Science, 7-1-26 Minatojima-minami-machi, Chuo-ku, Kobe, 650-0047, Hyogo, Japan\\
$^3$Centre for Astrophysics Research, University of Hertfordshire, College Lane, Hatfield, Hertfordshire AL10 9AA, UK\\
$^4$Advanced Research Computing, University College London, Judd St, London WC1H 9NE, UK
}

\date{Accepted XXX. Received YYY; in original form ZZZ}

\pubyear{2021}

\begin{document}
\label{firstpage}
\pagerange{\pageref{firstpage}--\pageref{lastpage}}
\maketitle

\begin{abstract}
Modelling star formation and resolving individual stars in numerical simulations of molecular clouds and galaxies is highly challenging.
Simulations on very small scales can be sufficiently well resolved to consistently follow the formation of individual stars, whilst on larger scales sinks that have masses sufficient to fully sample the IMF can be converted into realistic stellar populations. 
However, as yet, these methods do not work for intermediate scale resolutions whereby sinks are more massive compared to individual stars but do not fully sample the IMF.
In this paper, we introduce the \textit{grouped} star formation prescription, whereby sinks are first grouped according to their positions, velocities, and ages, then stars are formed by sampling the IMF using the mass of the groups. 
We test our \textit{grouped} star formation method in simulations of various physical scales, from sub-parsec to kilo-parsec, and from static isolated clouds to colliding clouds. 
With suitable grouping parameters, this star formation prescription can form stars that follow the IMF and approximately mimic the original stellar distribution and velocity dispersion. 
Each group has properties that are consistent with a star-forming region. 
We show that our \textit{grouped} star formation prescription is robust and can be adapted in simulations with varying physical scales and resolution.
Such methods are likely to become more important as galactic or even cosmological scale simulations begin to probe sub-parsec scales.
\end{abstract}


\section{Introduction}

Sink particles, first introduced by \cite{bate_modelling_1995} and subsequently developed further by \cite{hubber_sink_2013} and \cite{bleuler_towards_2014}, are used to replace high density regions in simulations, as the individual timesteps of the high density gas particles become too small to follow. 
Sink particles are as such commonly used to represent the sites of star formation in simulations from individual stars or dense cores \citep[e.g.][]{bate_formation_2003,price_magnetic_2008,bate_cluster_2009,bate_cluster_2012,lomax_prestellar_2015,boss_protostar_2019}, small groups of stars \citep[e.g.][]{girichidis_initial_2011,federrath_sfr_2012,balfour_star_2015,bertellimotta_imf_2016,ali_feedback_2019,he_clusters_2019,ntormousi_filaments_2019,dobbs_ymc_2020,liow_collision_2020,dobbs_magnetic_YMC_2021}, up to whole clusters \citep[e.g.][]{vazquezsemadeni_molecular_2007,renaud_antennae_2015,howard_YMC_2018,bending_photoionisation_2020,ali_hiiregion_2021}.
By replacing high density regions with sink particles, this method ensures a cheaper and more efficient computation. 
Sink particles are usually decoupled from the physics of hydrodynamics and only interact with each other and gas particles via gravity.
They can be used to incorporate physics such as following
accretion onto forming protostars \citep[e.g.][]{bate_modelling_1995, bonnell_accretion_1997,federrath_modeling_2010,dale_accretion_2011,padoan_turbulence_2011,stamatellos_accretion_2012,latif_black_holes_2015,jones_radiative_2018,kuznetsova_protostellar_2020}, and disc formation \citep[e.g.][]{stacy_multiples_2010,greif_protostellar_2012,hennebelle_disc_2020}.

In some smaller length scale and high mass resolution (i.e. low mass per gas particle) simulations \citep[e.g.][]{bate_cluster_2009,bate_cluster_2012,lomax_prestellar_2015}, sink particles are formed when fragmentation is resolved at the opacity limit \citep{Masunaga_2000}, so these sink particles are well resolved as individual stars. 
Nonetheless, most hydrodynamical simulations do not have such high mass resolution, so sink particles are often introduced when the Jeans mass criterion is still satisfied \citep{bate_resolution_1997}, which is usually at much lower densities than the opacity limit for fragmentation. 
This means that even though each sink is a self-collapsing star-forming region, they represent small groups of stars and not individual protostars. Thus, cluster evolution cannot be studied as well compared to using simulations with resolved star particles \citep[e.g.][]{bate_cluster_2012, fujii_initial_2015}.
This scenario also assumes that any stars that are formed are always bound within their parent sinks, which is commonly assumed in many stellar feedback simulations \citep[e.g.][]{dale_feedback_2014,geen_feedback_2016,geen_star_formation_2018,bending_photoionisation_2020}.
This situation would not capture, for example, the mass segregation of massive stars towards the centre of the larger clusters, or expulsion of stars due to close interactions, which may affect the evolution of clusters.

Instead of resolving sinks as stars, a workaround is to form star particles either using a local star formation efficiency function \citep{fujii_initial_2015} or a pre-determined initial mass function \citep[IMF;][]{hu_interstellar_2017,lahen_griffin_2020,ballone_stars_2021,hirai_sirius1_2021,hislop_dwarf_2021, smith_sensitivity_2021}.
This allows the introduction of star particles in lower resolution simulations. 
Nonetheless, mass is unable to be conserved locally in \cite{fujii_initial_2015} and \cite{smith_sensitivity_2021}, so this can affect the stellar distribution and thus the modelling of stellar feedback, if included, would be inaccurate \citep{dinnbier_feedback_2020}. 
Even though the method by \cite{ballone_stars_2021} is excellent in conserving global and local stellar properties, it is not developed to form stars dynamically in hydrodynamical simulations and therefore unable to model the dense gas distribution. 
The star formation method by \cite{hirai_sirius1_2021} that uses density-independent formulation of smoothed particle hydrodynamics (SPH) for gas hydrodynamics \citep{saitoh_disph_2013} can be applied to simulations of various length scales, however it can create gas with unequal masses, which is by design not permitted in some codes that implement standard SPH \citep[e.g.][]{price_gravity_2007, price_phantom_2018}.
Another way to introduce stars in the simulation is via ‘sink-star’ hybrid particles as demonstrated in
\cite{grudic_starforge_2021}, whereby a protostar is embedded in each sink and only interacts with the simulation through accretion from its sink reservoir and stellar feedback, but this method does not resolve stellar dynamics within the sinks.

A novel way to create stars from sinks in low mass resolution simulations was introduced by \cite{wall_flash_2019} and is hereafter referred to as \textit{single} star formation. 
Sink particles are formed using the prescription by \cite{federrath_modeling_2010}, then each sink particle forms stars that are sampled from the user-input IMF using Poisson sampling \citep{sormani_sinks_stars_2017}. 
The stellar population from each sink is self-consistent, i.e. each stellar population follows the IMF. 
This method of star formation ensures that stellar dynamical interaction happens not only between stars formed within the same sink but also between stars formed in different sinks. 
Moreover, local mass conservation is obeyed and the star formation prescription causes minimum disruption to the gas evolution code.
Nonetheless, to sample a complete IMF, each sink is expected to have $\gtrsim 10^2$ \msun{} in mass \citep{wall_flash_2019,smith_sensitivity_2021,rieder_ekster_2021}, which is usually assumed to be a `cluster sink' and only achievable in larger cloud or (sub-)galactic simulations, or simulations with very low mass resolution. 
As we investigate in this paper, applying the \textit{single} star formation method on parsec-scale or smaller simulations, in which sinks are usually assumed to be small groups of stars, results in oversampling of low mass stars and undersampling of high mass stars. 
These simulations usually have a mass per gas particle of about $10^{-1} - 10^{-4}$ \msun{}. Consequently, no sink is massive enough to sample a complete IMF, while the sinks are not small enough to be resolved as individual star either.

In this paper, we extend the method of converting sinks to stars as presented in \cite{wall_flash_2019} and \cite{rieder_ekster_2021} to apply to higher mass resolution simulations if sinks cannot be resolved as individual stars, but are individually not large enough to sample a complete IMF. 
We perform \textit{grouped} star formation, i.e. we group the sink particles and use the group mass to sample the IMF. 
We first group the sink particles according to certain length, speed, and time scales so that each group is approximately a consistent star-forming region. 
After sampling the IMF using the group mass, stars are placed within the sinks according to the sink masses to optimise the local conservation of mass. 
This method allows lower mass sinks to be grouped so that the group mass is large enough to sample the IMF robustly. 
This means that we can afford the formation of smaller size sinks so that the stellar distribution can mimic the dense gas distribution more accurately. 
The \textit{grouped} star formation method also allows us to test the limit of the \textit{single} star formation method.

This paper is sectioned as follow: in Section \ref{sec:numeticalmethods}, we lay out the prescription for \textit{grouped} star formation, and the simulation setup to study this method in different length scales.
The results are shown in Section \ref{sec:results}. We explore the effect of changing the random seed for star formation and varying the upper star mass limit of the IMF in Section \ref{sec:discussion}. Lastly, we summarise the paper in Section \ref{sec:conclusion}.

\section{Numerical methods}
\label{sec:numeticalmethods}

\subsection{Methods for group assignment and star formation}
\label{ssec:starformation}

In general, the \textit{grouped} star-forming prescription is divided into two steps: the grouping of sink particles and 
the conversion of sink particle mass into star particles. 
The group assignment occurs when new sink particles are formed, while the conversion of sink mass to stars happens whenever there are sink particles in the simulation.
The newly formed sinks are first arranged according to their masses in descending order. 
When the first sink particle is formed in the simulation, it forms a group by itself, since no group existed prior to that. 
After the first sink, for the $a$-th newly formed sink particle, we loop through all existing groups and add the new sink to a pre-existing sink group $g$, if:

\begin{enumerate}
    \item sink particle $a$ is within a distance $d_g$ away from the group $g$'s centre-of-mass (COM), i.e. $|\bm{r}_a - \bm{r}_{\textnormal{COM},g}| \leq d_g$, where $\bm{r}$ is the position vector, 
    
    \item sink particle $a$ is within a speed $v_g$ of the group $g$'s centre-of-mass velocity, i.e. $|\bm{v}_a - \bm{v}_{\textnormal{COM},g}| \leq v_g$, where $\bm{v}$ is the velocity vector, 
    
    \item the creation time of sink particle $a$ is at most $\tau_g$ greater than the oldest member in group $g$, and
    
    \item the sink particle $a$ is the most bound to group $g$ compared to the other groups.
\end{enumerate}

\noindent For ease of reference, henceforth the first three conditions are called the distance, speed, and age criteria respectively. 
These criteria are set so that each group is approximated as a self-consistent star-forming region, such that $d\_g$, $v\_g$, and $\tau\_g$ are the length, speed, and time scales of a typical star-forming region respectively. 
The speed criterion ensures that any sinks that are likely moving away from the group are not included within the group, while the age criterion makes sure that sinks formed at later times are not grouped together with sinks formed much earlier, even though they could coincidentally be located near each other.
The fourth condition is not essential but necessary in case a sink particle satisfies the grouping criteria of more than one group.
If the $a$-th sink particle fails to satisfy all of the above conditions, then it will create a new group by itself, and the process continues for the other sink particles. 
Each sink particle is assigned to only one group and will be in the same group in subsequent timesteps to keep track of the stellar population in each group. 

After the group assignment, a population of stars is introduced for each group of sink particles.
The introduction of stars is similar to the \textit{single} star formation method used in \cite{wall_flash_2019} and \cite{rieder_ekster_2021}, but instead of taking individual sink mass to sample the whole IMF, we take the total sink mass in the group to do that. 
In our simulations, we use the Kroupa IMF \citep{kroupa_2001} to generate a list of stars with a user-defined mass range (a typical choice would be from 10$^{-2}$ to 10$^2$ \msun{}). 
Next, we use the sink particle masses to form a probability distribution list, such that stars are more likely to be allocated at higher mass sink particles. 
Lastly, each star is given a position $\textbf{r}\_{star}$ and velocity $\textbf{v}\_{star}$ such that

\begin{equation}
\begin{aligned}
    \textbf{r}\_{star} &= \textbf{r}\_{sink} +  \textbf{r}\_{random}, \\
    \textbf{v}\_{star} &= \textbf{v}\_{sink} + \textbf{v}\_{gaussian},
\end{aligned}
\end{equation}

\noindent where $\textbf{r}\_{sink}$ and $\textbf{v}\_{sink}$ are the position and velocity of the allocated sink particle.
$\textbf{r}\_{random}$ is the randomly-assigned position within the sink particle accretion radius $r\_{acc}$, i.e. $0 \leq r\_{random} \leq r\_{acc}$ to simulate star formation in dense regions, while $\textbf{v}\_{gaussian}$ is the velocity sampled from a Gaussian distribution with standard deviation equal to the local velocity dispersion of the original gas particle. 
It is difficult to assign the true underlying gas velocity to the stellar velocity at birth, as the underlying gas particles are accreted onto sinks before stars are formed, so the velocity field would have been different.
Nonetheless, the velocity of stars at birth closely mimics the approximated underlying gas velocity field created using other gas particles, as the median difference in velocity magnitude between stellar velocity and the underlying gas velocity field is $\lesssim$ 10\%, and the median difference in angle is $\lesssim \pi/10$ for our models.
Our choice of using a Gaussian distribution to add on to $\textbf{v}\_{sink}$ could potentially be modified to include a power law tail to model the inclusion of runaway stars \citep{perets_runaways_2012}, however we are not primarily studying such stars here, and this would involve a further arbitrary choice of power law tail and cut off. We note that stars can still be ejected dynamically through interactions as the cluster evolves.
Our choice of position and velocity assignment also ensures that most stars are evolved close to their parent sinks.


For each group, the sink particle masses are reduced according to the total star mass located within them. 
However, although unlikely, the total stellar mass can be greater than the mass of the sink the stars are located in, due to the probabilistic nature of allocating stars. 
To overcome the issue of having negative sink mass, the sink mass is set to $m\_{thres}$ (here, $m\_{thres}$ is chosen as the mass of a gas particle) when the difference in total star mass and their parent sink is less than $m\_{thres}$. 
This mass difference is accumulated for all sinks in the group and is then reduced from all sinks in the group in proportion to their masses. 
The mass conservation of the group is guaranteed using this scheme. Lastly, the sink particles shrink in size to keep their creation density constant. 
In subsequent timesteps, when a new sink joins an existing group, its mass is added to the leftover mass of the sink group. The new total group mass is then used to sample the IMF and make new stars, and finally, the masses of the sinks in the group are reduced as described.

\subsection{Simulation setup}
\label{ssec:simulation}

We use \texttt{Ekster} \citep{rieder_ekster_2021,ekster_code} to perform our simulations. 
\texttt{Ekster} is a multiphysical code that combines gas hydrodynamics, gravitational dynamics and stellar evolution via the \texttt{AMUSE} interface \citep{amuse_2018}. 
The gravitational dynamics between gas and non-gas (sinks and stars) are coupled using \texttt{Bridge} \citep{fujii_bridge_2007}.
For gas hydrodynamics, we use \texttt{Phantom} \citep{price_phantom_2018}, a smoothed particle hydrodynamical (SPH) code for astrophysics. 
Artificial viscosity is included to model shocks \citep{monaghan_1997}. 
The standard values of artificial viscosity parameters $\alpha\_{min}^{\textnormal{AV}} = 0.1$, $\alpha\_{max}^{\textnormal{AV}} = 1$, and $\beta^{\textnormal{AV}} = 2$ are used \citep{morris_1997} except for Section~\ref{sssec:icl20} where higher values are used.
\texttt{PeTar} \citep{wang_petar_2020} is used for the fast calculation of gravitational dynamics between and among sink particles and stars. 
Lastly, we use \texttt{SeBa} \citep{portegies_zwart_seba1_1996}, a parametric code to calculate stellar evolution. Sink particles are introduced to replace high density gas particles, following the prescription used in \cite{bate_modelling_1995} and \cite{price_phantom_2018}, when the gas density exceeds the sink creation density $\rho\_{sink}$. 
Depending on the system studied, we adjust $\rho\_{sink}$ such that the minimum Jeans mass criterion is satisfied \citep{bate_resolution_1997}. 
Sinks are allowed to accrete mass.

\subsection{Initial conditions}

\begin{figure}
    \centering
    \includegraphics[width=\columnwidth]{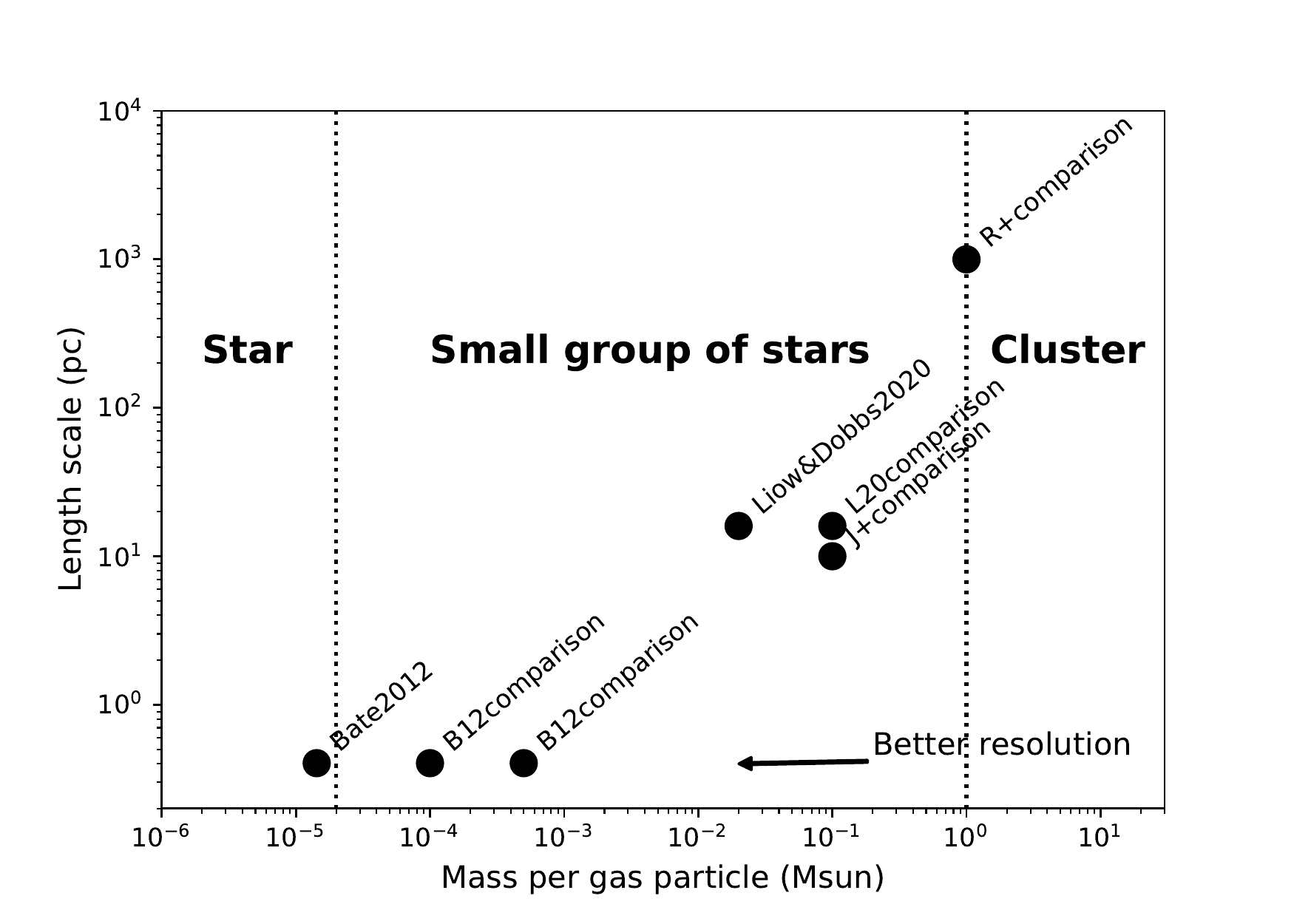}
    \caption{The relationship of length scale and mass resolution is shown for the original simulations and models compared in this paper (see the text for details of different simulations presented). 
    The plot is roughly divided into three regions by the mass thresholds $2\times10^{-5}$ \msun and 1 \msun{} which show the nature of sink particle (bold text) in the simulations. 
    Resolution is traditionally defined as the inverse of mass per gas particle.}
    \label{fig:length_res}
\end{figure}

We test the \textit{grouped} star formation prescription on simulations of different length scales and mass resolutions as shown in Figure \ref{fig:length_res}. 
The aim is to demonstrate that \textit{grouped} star formation method is needed at the mass resolution regime in which each sink is usually considered as a small group of stars. 
In each comparison described below, we vary the grouping parameters as described in Section \ref{ssec:starformation}. 
The `no grouping' case, i.e. $d\_g = v\_g = \tau\_g = 0$, is equivalent to the \textit{single} star formation prescription used in \cite{wall_flash_2019} and \cite{rieder_ekster_2021}. 
On the other hand, the `all grouping' case simply groups all sinks as one, and we achieve this by setting $d\_g$, $v\_g$ and $\tau\_g$ unphysically large, usually $d\_g = 1000$ pc, $v\_g = 1000$ km/s, and $\tau\_g = t\_{ff}$ if the period of interest is within a free-fall time, or $\tau\_g = 1000 \ t\_{ff}$ otherwise. 
The main other settings that we explore are the `standard' case, i.e. $d\_g = 1$ pc, $v\_g$ = 1 km/s, and $\tau\_g=t\_{ff}$, where $t\_{ff}$ is the free-fall timescale of the system, and the `turbulent' case, i.e. the same as the`standard' case except with $v\_g$ equals to the initial turbulent speed (coincidentally, it is approximately 3 km/s in most of our different comparisons).
The standard case is chosen as such because they are convenient and appropriate choices for the length and speed scales of a typical star-forming region. 
Besides, the mass estimate for the group, in this case, $\sim d\_g v\_g^2/G \approx 230$ \msun{} turns out to be the same order of magnitude of the mass expected to sample a complete IMF \citep{wall_flash_2019, smith_sensitivity_2021}.
We also test other grouping parameters specific to individual comparisons and they are explained in detail in the subsequent subsections. 
We list the different models we use and the different grouping parameters for each in Table \ref{tab:initialconditions}.

\subsubsection{Sub-parsec-scale isolated cluster simulation}
\label{sssec:icb12}

We first test our sink method on a simulation of cluster formation in a 500 \msun{} isolated cloud. 
We use the simulation of \cite{bate_cluster_2012} who perform a 35 million particle (mass per gas particle $m\_{SPH}\approx 1.4 \times 10^{-5}$ \msun{}) simulation which is able to resolve star formation down to a mass of $\sim 10^{-2}$ \msun{}. 
We choose this simulation, which includes radiative transfer to model stellar heating \citep{whitehouse_sph_radiation_2005}, rather than \cite{bate_cluster_2009}, even though we don't include this additional physics. 
This is partly because the data was readily available, and also because with the radiative feedback, the IMF produced is much closer to a Kroupa IMF which matches our choice of IMF. 
We describe the initial conditions below. 
The aim of our simulations is to reproduce the properties of the stars in the original simulation of \cite{bate_cluster_2012}, but using a gas resolution much lower than the original simulation. 
We then compare the results from \cite{bate_cluster_2012} with our lower resolution simulations performed with Ekster. 
Our initial conditions are the same except for the number of particles. 
The cloud has a mass of 500 \msun{} and a radius of 0.404 pc initially, which gives an initial cloud density of $1.2 \times 10^{-19}$ \gcm{} and a spherical free-fall time of 0.19 Myr. 
We apply exactly the same velocity field as \cite{bate_cluster_2012}, which has a velocity dispersion of $\approx 3$ km/s. 

\cite{bate_cluster_2012} started with an initial temperature of 10.3 K and used SPH with radiative transfer in the flux-limited diffusion approximation \citep{whitehouse_sph_radiation_2005} in their simulation. 
As their sink particles are formed during the second collapse of protostellar cores \citep{Masunaga_2000}, they are well resolved as stars and brown dwarfs. 
On the other hand, we simply adopt the isothermal equation of state at a temperature of 10 K as our chosen $\rho\_{sink} = 10^{-16}$ \gcm{} is within the first phase of protostellar collapse, which can be modelled as isothermal collapse \citep{masunaga_collapse_1998, Masunaga_2000}. 
Our resolution is $N\_{SPH} = 5 \times 10^6$ ($m\_{SPH}=10^{-4}$ \msun{}) as opposed to $N\_{SPH} = 3.5 \times 10^7$ ($m\_{SPH} \approx 1.4 \times 10^{-5}$ \msun{}) used in \cite{bate_cluster_2012}, however our resolution and $\rho\_{sink}$ allow us to resolve the Jeans mass criterion \citep{bate_resolution_1997}. 
Unless otherwise stated, our gravitational softening length $\epsilon$ in this comparison is set at 0.001 pc $\approx 200$ AU.
In \cite{bate_cluster_2012}, gravitational softening between sink particles is turned off. 
The initial accretion radius of sink particles $r\_{acc} = 0.001$ pc, the same order of magnitude of the Jeans length in this system.

Based on \cite{bate_cluster_2012}, we set the grouping parameters described in \ref{ssec:starformation} as follows. 
For the `standard' case, $d\_g = 1$ pc and $v\_g = 1$ km/s are chosen as described.
For the `turbulent' case, the grouping parameters are the same as the `standard' case except for $v\_g = 3$ km/s, which is approximately the turbulent velocity dispersion of this gas cloud.
The \textit{single} star formation method is exactly the same as setting both $d\_g$ and $v\_g$ to zero, i.e. the `no grouping' case, as these settings prevent the formation of any groups at all.
We also consider the opposite extreme, i.e. the `all grouping' case by setting both $d\_g$ and $v\_g$ to unphysically large values of 1000 pc and 1000 km/s respectively to group all the sinks into one single group.

We include the age criterion $\tau\_g = t\_{ff}$ but in this comparison, this check is irrelevant as we only compare up to the free-fall time, so the sink particles and stars in all our simulations are definitely formed within that timescale. 
Lastly, we also perform additional simulations with lower resolution and with no gravitational softening separately. 
The `low resolution' model has $N\_{SPH}=10^6$ ($m\_{SPH} = 5 \times 10^{-4}$ \msun{}).
This model is used to check whether our star formation method works in a lower resolution of the same length scale. 
For the `no softening' model, the gravitational softening length $\epsilon$ is turned off but the gas particle softening length remains at 0.001 pc, the same as in other simulations in this comparison. 
This setting of $\epsilon <$ gas smoothing length is cautioned by \citet{bate_resolution_1997} as fragmentation at scales less than the gas smoothing length can be artificially induced.
However, we run this model to investigate the dynamics of stars formed from our star formation prescription. 
The details of the models for this comparison are listed in the upper section of Table \ref{tab:initialconditions}.

For star formation, we sample stellar masses from the Kroupa IMF \citep{kroupa_2001}. 
In Section \ref{ssec:resultb12} we take a mass range from 0.01 \msun{} to 100 \msun{} which extends down to the minimum mass of stars resolved in \cite{bate_cluster_2012}.
However since 100 \msun{} is large compared to the typical group size in the models in this comparison, in Section \ref{ssec:varying_limit} we also test taking a mass range from 0.01 \msun{} to the individual mass of the group. 

\subsubsection{Parsec-scale cloud-cloud collision simulation}
\label{sssec:icl20}

A cloud-cloud collision simulation from \cite{liow_collision_2020} is chosen to test our star formation method on larger scales with higher sink masses. 
In this case, individual stars are not resolved, rather each sink is simply a small group of stars. 
Another difference compared to \cite{bate_cluster_2012} is that in this larger scale simulation, sinks form at much wider distances from each other. 
The initial conditions are described in \cite{liow_collision_2020} (known as the model with low collision speed, standard density, and low turbulence) but we also describe them again below. 

Two ellipsoidal clouds of mass $5\times 10^4$ \msun{} and minor radii of 7 pc collide along the major radius of 16 pc. 
This gives an initial cloud density of $1.03 \times 10^{-21}$ \gcm{} and a spherical free-fall time of 2.07 Myr. 
In the original simulation, the resolution is $5 \times 10^6$ ($m\_{SPH} = 0.02$ \msun{}), however the resolution we use is $10^6$ SPH particles ($m\_{SPH} = 0.1$ \msun{}). 
The gas clouds are initially about 0.3 pc apart, and are subjected to two separate turbulent fields. 
In this setup, the turbulence has a velocity dispersion of 2.5 km/s, and the clouds are given a relative velocity of 10 km/s. 
We choose to compare our models at $t_{10\%}$, the time when 10\% of the gas mass is converted to sink particles, as the sink particle distribution shows multiple star-forming regions, notably the central cluster at the collision site, and the filamentary structures that are in the parts of the clouds yet to collide.
The ongoing collision also gives multiple velocity components to be considered. 
The external forces exerted by the colliding clouds cancel out at the collision site, giving a zero net momentum for the star-forming regions at the collision site, whilst sinks away from the area where the clouds are colliding are still moving at about 10 km/s relative to each other. 

In the original model by \cite{liow_collision_2020}, sinks are formed at $\rho\_{sink} = 10^{-18}$ \gcm{} which was typically too low for them to be considered individual stars but enough to satisfy the Jeans mass criterion \citep{bate_resolution_1997}.
Here, we use the same $\rho\_{sink}$ as our sink creation density.
In this comparison, we also set the artificial viscosity parameter $\beta^{\textnormal{AV}} = 4$ as for the original simulation to minimise the effect of particle penetration in high speed shocks \citep{price_comparison_2010}.
Similar to the original model, the gravitational softening length and the initial accretion radius $r\_{acc}$ are set to 0.01 pc. 
We consider the `no grouping', `standard', `turbulent' (with $v\_g = 3$ km/s, approximately the initial turbulent speed of the clouds), and `all grouping' cases, similar to the comparison described in Section \ref{sssec:icb12}. 
The details of the models in this comparison are listed in the second part of Table \ref{tab:initialconditions}.

In this comparison, we choose a mass range of 0.5 - 100 \msun{}, as it roughly reflects the mass distribution of sinks from the original simulation at $t_{10\%}$, which is also our chosen timescale for comparison. 
Unfortunately, unlike the models described in Section \ref{sssec:icb12}, we cannot resolve the full IMF with larger scale simulations, so we simply compare our models with lower resolution to the original simulation. 
We could, however, potentially choose a lower stellar mass limit for our model (see Buckner et al. in prep which has a lower stellar mass limit of 0.01 \msun{}).

\subsubsection{Parsec-scale isolated cluster simulation}
\label{sssec:icjaffa}

We adopt the \textit{grouped} star formation method to a parsec-scale, gravitational collapse of an isolated cloud simulation originally performed by \cite{jaffa_sims_prep}. 
We do not rerun the simulation, rather here we post-process the results. 
The mass and radius of the cloud are $10^4$ \msun{} and 10 pc respectively, which gives an initial cloud density of $1.61\times 10^{-22}$ \gcm{} and a free-fall time of 5.24 Myr. 
The resolution used in the simulation is $10^6$ SPH particles, which corresponds to $m\_{SPH} = 0.01$ \msun{}.

The cloud has an initially uniform density and a turbulent velocity field, with a natural mix of compressive and solenoidal modes and a power spectrum of $P(k) \propto k^{-4}$.
The turbulence is allowed to decay as the cloud collapses to form filamentary structures, which merge to feed a central cluster with some smaller structures around it. 
Sink particles are inserted when gas reaches a density of $10^{-18}$ \gcm{}, with some further checks such as proximity to other sinks \citep[see][]{hubber_gandalf_2018}. 
The gravitational softening applied to interactions between two sink particles is the mean of their sink radii, which are set to their SPH smoothing length, $h \sim 10^{-2}$ pc. 
Sinks continue to accrete but are not allowed to merge, and by 10 Myr more than 60\% of the gas has been turned into sink particles.
The simulation is terminated at 20 Myr, but we note that some important physical processes such as feedback are not modelled in this simulation so the later evolution of the cloud and cluster may be affected by this. 
The full details of the simulation can be found in \cite{jaffa_sims_prep}. 

We use this simulation partly to test the grouping age criteria. 
In the simulations described so far in Sections \ref{sssec:icb12} and \ref{sssec:icl20}, the age criteria is not relevant because the stars form in a short timescale, within a free-fall time. 
This simulation however is run for 20 Myr $\approx 3.5 t\_{ff}$, and so provides a good test of the age criteria. 
Although the resolution is not that high, the 20 Myr timescale means the calculation still required considerable computational resources to run, and so we do not repeat the calculation with different grouping parameters, rather we apply these post process. 
Therefore, instead of forming stars dynamically, we apply a modified \textit{grouped} star formation prescription on the sinks from the original simulation at specific snapshots to form stars, keeping track of the group indices assigned to the groups at previous snapshots. 
Unfortunately, this means that we cannot analyse the velocity dispersion or the stellar distribution, since we do not follow the dynamics, but we can still consider the mass functions and properties of the groups. 
Similar to our previous comparisons, we choose models with different grouping parameters to represent `no grouping', `standard', `turbulent' ($v\_g = 3$ km/s, approximately the turbulent velocity dispersion of the cloud), and `all grouping' cases. Here, the `all grouping' case sets all three grouping parameters ($d\_g, v\_g$ and $\tau\_g$) unphysically large. 
To investigate the age criteria more robustly, we include the `no age check' model, whereby the grouping parameters are the same as the `standard' model except that $\tau\_g$ is set unphysically high to prevent triggering this condition check. 
We also include the `free-fall' model, whereby the grouping parameters are the same as the `all grouping' model except that $\tau\_g = t\_{ff} = 5.24$ Myr, since our comparison is up to 20 Myr ($\approx 3.5 t\_{ff}$). 
In this comparison, we use the Kroupa IMF with mass range 0.5 - 100 \msun{}. 
The parameters of the models are listed in the third part of Table \ref{tab:initialconditions}.

\subsubsection{Kiloparsec-scale spiral arm simulation}
\label{sssec:icrieder}

For completeness, we take the stars from the spiral arm simulation by \cite{rieder_ekster_2021} and study the mass functions. 
This simulation models star cluster formation along a section of spiral arm. 
The simulation \citep[labelled standard-arm-1 in][]{rieder_ekster_2021} models a region of dimensions 600 pc, which includes a number of molecular clouds. 
The mass resolution is $m\_{SPH}=1$ \msun{}.

We do not rerun this simulation because in the original simulation, \cite{rieder_ekster_2021} uses the \textit{single} star formation method to form stars. 
If the IMF of the stars in the original simulation is already close to a Kroupa IMF, then we can assume that the \textit{single} star formation method is suitable for simulations of this length scale and mass resolution, or larger. 
In the original simulation, the Kroupa IMF is used as the input IMF, and the chosen mass range is 0.1 to 100 \msun{}. 

\begin{table*}
    \centering
    \begin{tabular}{l|l|l|l|l|l|l|l|l|l|l|l}
    \hline \hline
    Model & $d\_{g}$ & $v\_{g}$ & $\tau\_g$ & $N\_{SPH}$ & $\epsilon$ & $N\_{sinks,g}$ & $N\_{stars,g}$ & $\Tilde{m}\_{stars,g}$ & $M\_{sinks}$ & $M\_{stars}$ & $f$  \\ 
    
     & (pc) & (km/s) & $t\_{ff}$ & ($10^6$)  & (pc) & & & (\msun{}) & (\msun{}) & (\msun{}) &  \\
    \hline \hline
    B12              & -    & -    & - & 35 & 0     &  -    & -  & - & 19.19 & -     &-     \\
    B12NoGrouping    & 0    & 0    & 0 & 5  & 0.001 & 1126 & 69 & 0.02 & 21.59 & 1.13  & 0.050  \\
    B12Standard      & 1    & 1    & 1 & 5  & 0.001 & 41   & 27 & 0.39 & 11.09 & 11.01 & 0.498  \\
    B12Turbulent     & 1    & 3    & 1 & 5  & 0.001 & 28   & 21 & 0.45 & 10.72 & 14.61 & 0.576  \\
    B12AllGrouping   & 1000 & 1000 & 1 & 5  & 0.001 & 1    & 1  & 17.30 & 5.28  & 17.30 & 0.766 \\
    B12NoSoftening   & 1000 & 1000 & 1 & 5  & 0     & 1    & 1  & 13.70 & 9.23   & 13.70 & 0.597  \\
    B12LowResolution & 1000 & 1000 & 1 & 1  & 0.001 & 1    & 1  & 23.16 & 0.30   & 23.16 & 0.987  \\
    \hline
    L20            & -    & -    & - & 5  & 0.01 & -    & -  & - & $1.01 \times 10^4$ & -     & -     \\
    L20NoGrouping  & 0    & 0    & 0 & 1  & 0.01 & 1817 & 1742 & 2.68 & $1.01 \times 10^3$ & $4.91 \times 10^3$  & 0.708 \\
    L20Standard    & 1    & 1    & 1 & 1  & 0.01 & 808  & 760  & 4.84 & $1.33 \times 10^3$ & $5.50 \times 10^3$  & 0.804 \\
    L20Turbulent   & 1    & 3    & 1 & 1  & 0.01 & 190  & 187  & 21.62 & $7.11 \times 10^2$ & $6.49 \times 10^3$  & 0.901 \\
    L20AllGrouping & 1000 & 1000 & 1 & 1  & 0.01 & 1    & 1    & $6.39 \times 10^3$ & 0.34               & $6.39 \times 10^3$  & 0.999 \\
    \hline
    J+             & -    & -   & -    & 1  & $\sim 0.01$  & -   & -   & - & $8.32 \times 10^3$ & -     & -      \\
    J+NoGrouping  & 0    & 0    & 0    & -  & -    & 208 & 204 & 28.86 & - & $7.46 \times 10^3$  & 0.896  \\
    J+Standard    & 1    & 1    & 1    & -  & -    & 160 & 159 & 34.07 & - & $7.81 \times 10^3$  & 0.939  \\
    J+Turbulent   & 1    & 3    & 1    & -  & -    & 85  & 82  & 38.60 & - & $8.04 \times 10^3$  & 0.967  \\
    J+NoAgeCheck  & 1    & 1    & 1000 & -  & -    & 155 & 151 & 36.98 & - & $7.72 \times 10^3$  & 0.928  \\
    J+FreeFall    & 1000 & 1000 & 1    & -  & -    & 3   & 3   & $5.49 \times 10^2$ & - & $8.28 \times 10^3$  & 0.995  \\
    J+AllGrouping & 1000 & 1000 & 1000 & -  & -    & 1   & 1   & $8.32 \times 10^3$ & - & $8.32 \times 10^3$  & 0.999  \\
    \hline 
    R+ (1.80 Myr)  & 0    & 0   & 0    & $\approx 5$  & 0  & 198   & 198   & $2.75 \times 10^2$ & $7.59 \times 10^2$ & $6.05 \times 10^4$     & 0.988  \\
    R+ (2.40 Myr)  & 0    & 0   & 0    & $\approx 5$  & 0  & 713   & 713   & $2.39 \times 10^2$ & $2.62 \times 10^3$ & $1.93 \times 10^5$     & 0.987  \\
    \hline \hline
    \end{tabular}
    \caption{List of the models performed. The first three columns after the model names are the \textit{grouped} star formation parameters: grouping distance $d\_g$, grouping speed $v\_g$, and grouping age $\tau\_g$ in terms of free-fall time $t\_{ff}$. The next two columns are the number of SPH particles $N\_{SPH}$, and gravitational softening length $\epsilon$ .
    The subsequent columns show the statistics of the models: number of sink groups $N\_{sinks,g}$, number of star groups $N\_{stars,g}$, median star group mass $\Tilde{m}\_{stars,g}$, total sink mass after star formation $M\_{sinks}$, total star mass $M\_{stars}$, and star mass fraction $f$, i.e. the fraction of star mass over the total non-gas mass. The first part is the comparison with \protect\cite{bate_cluster_2012} at the free-fall time of 0.19 Myr, the second part is the comparison with \protect\cite{liow_collision_2020} at 1.75 Myr, while the third part is the comparison with \protect\cite{jaffa_sims_prep} at 20 Myr. For the last part, i.e. comparison with \protect\cite{rieder_ekster_2021}, the time considered is listed beside the models.  
    }
    \label{tab:initialconditions}
\end{table*}

\section{Results and Discussion}
\label{sec:results}

\subsection{Sub-parsec-scale isolated cluster simulation}
\label{ssec:resultb12}

\begin{figure}
    \centering
    \includegraphics[width=\columnwidth]{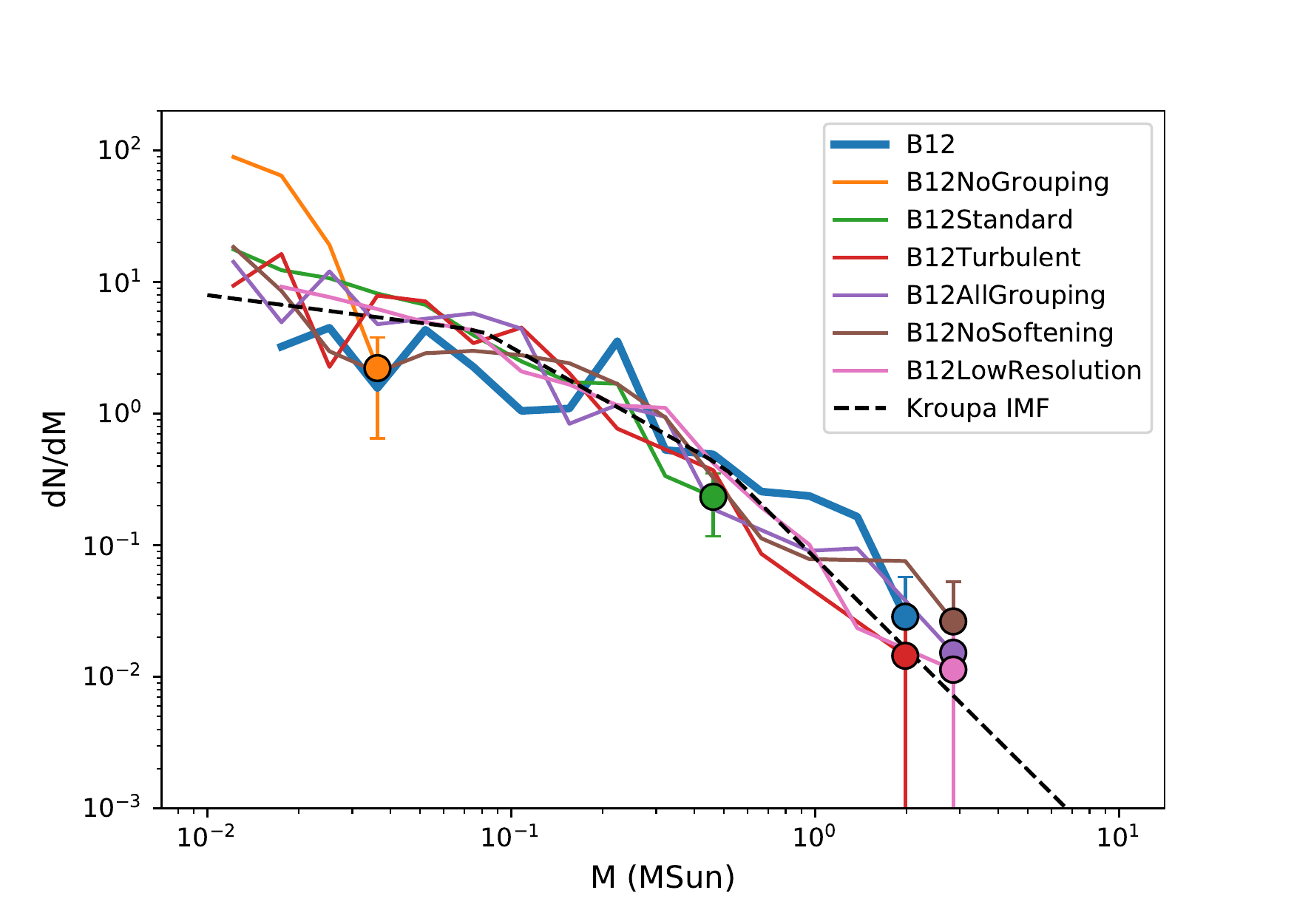}
    \caption{The IMF of our models are compared with that of \protect\cite{bate_cluster_2012}, labelled as B12, at the free-fall time of 0.19 Myr. 
    The markers are the maximum mass bin in the respective models.
    The errorbars are the standard deviations from Poisson sampling.
    The black dashed line is the analytical Kroupa IMF. 
    }
    \label{fig:B12IMF}
\end{figure}

Figure \ref{fig:B12IMF} shows the IMF of the models to compare with \cite{bate_cluster_2012}, i.e. model B12, at the free-fall time of 0.19 Myr. 
The IMF of B12 is similar to the Kroupa IMF. 
The `no grouping' model (orange) that uses the \textit{single} star formation to form stars is unable to sample a complete Kroupa IMF.
The average mass reservoir to form stars, i.e the average sink mass, is 21.59 \msun{}/1126 = 0.02 \msun{}, meaning not many stars above this threshold are expected to form, causing an oversampling of low-mass stars and undersampling of high mass stars. 
Moreover, as shown in Table \ref{tab:initialconditions}, the star mass fraction, defined as the fraction of star mass over the total of sink and star masses, is $f = 0.05$, so most of the mass in each sink is unable to be converted to stars.
The value of $f$ is low because for a population of stars to be introduced for each group (in this model, each group is a sink), the group mass has to exceed a mass threshold sampled from the Kroupa IMF. 
This is difficult for this model because the probability of forming a star less than 0.02 \msun{} is only 7\%. 
We discuss setting the upper star mass limit to be equal to the group mass as a potential improvement to produce a higher $f$ in Section \ref{ssec:varying_limit}.

With \textit{grouped} star formation of varying degrees, the maximum star mass bin (markers in Figure \ref{fig:B12IMF}) is greatly increased. 
The `standard' model (green) with $d\_g=1$ pc and $v\_g=1$ km/s is still unable to sample the full Kroupa IMF. 
We find that by setting $v\_g$ to 3 km/s, about the initial root-mean-square turbulent velocity of the cloud, while keeping the same $d\_g$, the `turbulent' model (red) can sample the full Kroupa IMF like the `all grouping' model (purple) that collects all sinks as one giant group, simply because setting the grouping parameters that match the length scale and turbulence of the initial cloud approximately covers the whole parameter space of this system. 
In all the models, by summing the mass of sinks and stars, we see that the total mass differs between models. 
However, the stellar mass in the `all grouping' case most closely matches that of the B12 model, whilst the `no grouping' model (\textit{single} star formation) massively underproduces stellar mass as compared to the original B12 calculation. 

We included two more models to test the effect of changing gravitational softening length and resolution on the IMF. 
The `no softening' model (brown) has similar IMF as the `all grouping' model simply because the total sink mass of the system is similar regardless of the stellar dynamics. 
Similarly, the `low resolution' model (pink) is also able to sample a complete Kroupa IMF like the `all grouping' model, showing the robustness of the \textit{grouped} star formation at different resolutions. 

\begin{figure}
    \centering
    \includegraphics[width=\columnwidth, trim={1cm 1cm 1cm 1cm}]{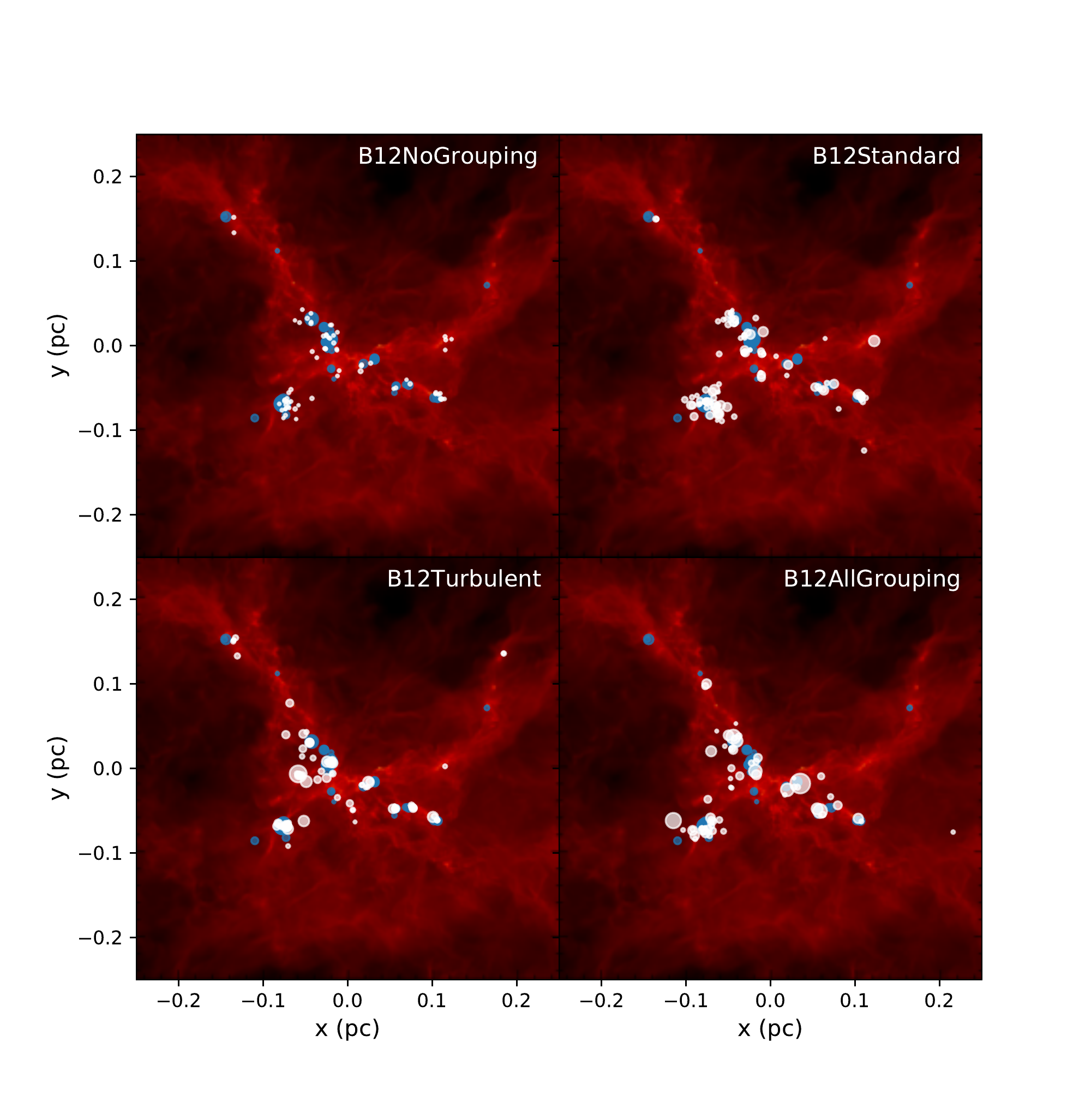}
    \caption{The stars from the `no grouping', `standard', `turbulent', and `all grouping' models (white) are compared against the sinks from \protect\cite{bate_cluster_2012} (blue) at the free-fall time of 0.19 Myr. 
    The marker radius is proportional to the particle mass. 
    The gas distribution from the original simulation is set as background for reference, as the gas distribution from the individual models are visually identical to the original one.
    }
    \label{fig:B12SD}
\end{figure}

Figure \ref{fig:B12SD} shows the stellar distributions of the `no grouping', `standard', `turbulent' and `all grouping' models. 
The stars in the `no grouping' model (top left), formed from the \textit{single} star formation method, are less visible due to their low mass. 
The same can be said for the `standard' model (top right), even though we can observe that the low mass stars are generally located at the region stars from B12 are expected. 
The `turbulent' model (bottom left) contains massive stars and reproduces the spatial stellar mass distribution reasonably well. 
Lastly, we note the `all grouping' model (bottom right) that groups all sinks together can form stars that are the most massive among all models, but the stellar distribution of this model deviates from that of B12 the most.
On scales of around $\sim 10^{-2}$ pc, mass is not conserved locally in this model. 
However, it is not necessarily important to model the exact spatial distribution of stars in a cluster in larger scale simulations where the minimum resolvable length scale may be $>>10^{-2}$ pc.
The `no softening' and `low resolution' models are not shown in Figure \ref{fig:B12SD}, however in terms of the degree of grouping, they are similar to the `all grouping' model but with different gravitational softening or resolution, so they are similar to the `all grouping' model in terms of their stellar distribution. 
In summary, the models that group the most sinks like the `turbulent' or `all grouping' models produce IMFs which match the IMF of B12. 
This is not surprising since the whole system is so small that it takes the whole region to sample the IMF. 
Thus, grouping most (which happens in the `turbulent' model where grouping parameters approximately equal to the scales of the cloud) or all the sinks together (which happens in the `all grouping' model) works best, and the \textit{single} star formation method is unsuitable. 
Grouping most or all the sinks together means that the spatial distribution matches that of B12 less well, but this is less concerning in simulations where the gas resolution is low as compared to the size of the clusters.

\begin{figure}
    \centering
    \includegraphics[width=\columnwidth]{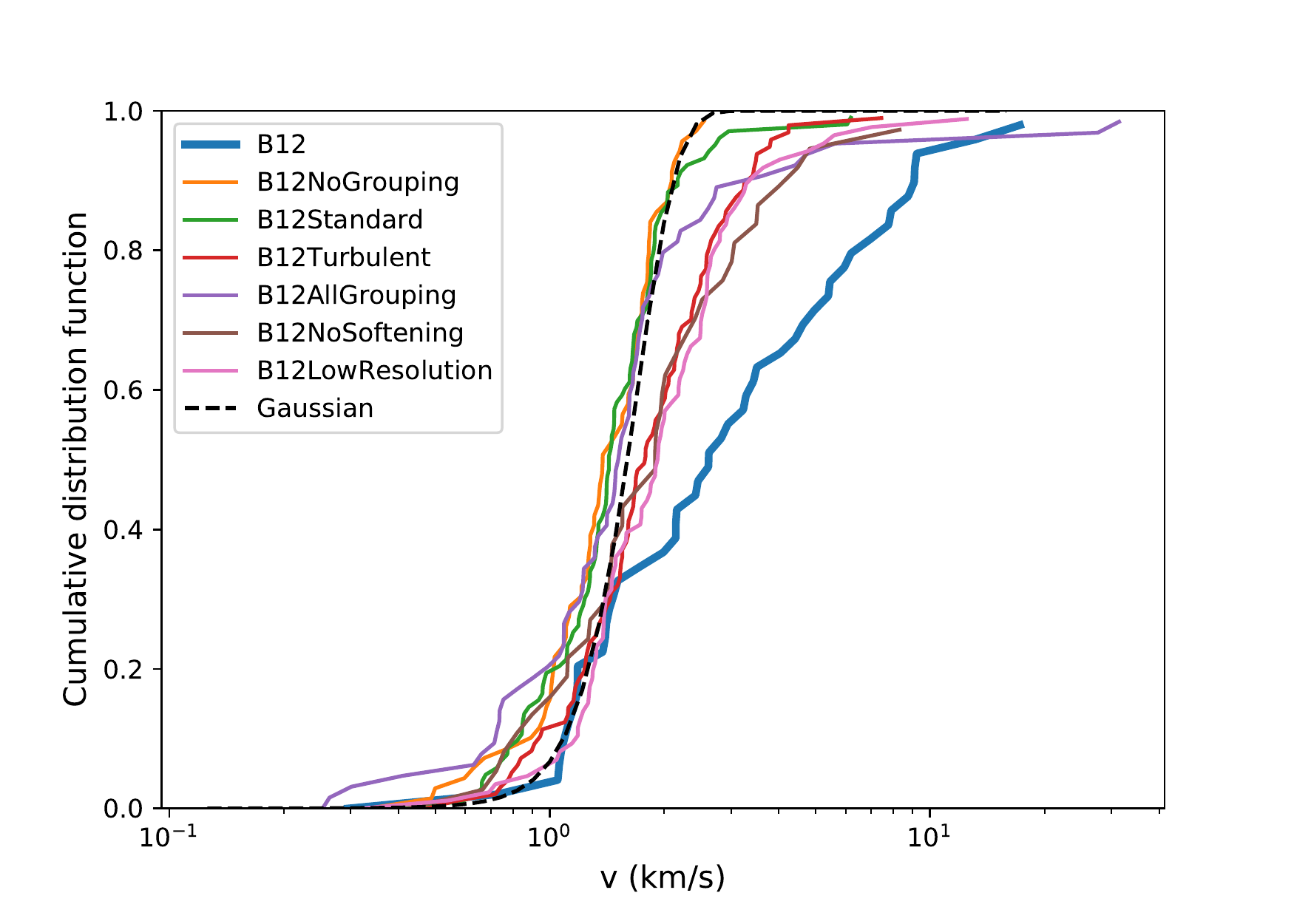}
    \caption{The cumulative velocity distribution of the models are shown are compared with that of \protect\cite{bate_cluster_2012} at 0.19 Myr. 
    The black dashed line is the Gaussian cumulative distribution function with mean 1.6 km/s and standard deviation of 0.4 km/s.
    It is not a fit and is included solely for comparison.}
    \label{fig:B12VD}
\end{figure}

Figure \ref{fig:B12VD} shows the cumulative distribution velocity function of the stars for all our models as compared to that of B12 (blue). 
We included a Gaussian distribution (black dashed line) with arbitrary mean of 1.6 km/s and standard deviation of 0.4 km/s to show that all our star formation models have velocity dispersion that is approximately Gaussian (with varying means and variances), as expected given how velocity is assigned to the stars \citep{wall_flash_2019}. 
The models with a greater degree of grouping have velocity dispersion closer to B12, but even with models that group all sinks as one like the `all grouping' model (purple), they are still visually different from the velocity dispersion of B12. 
We use the Kolmogorov-Smirnov (KS) test with permutation \citep{kstest_permutation_1995} to test the hypothesis of equal distribution between the velocity dispersion of our models with that of B12. 
With a significance level of 0.01, the result shows that all models except the `no softening' model (brown) have velocity dispersions that are different from that of B12. 
Statistically, this means that the stellar dynamics of the `no softening' model and the original simulation of B12 could be similar, which would likely be caused by the similar stellar mass distribution and the fact that both have gravitational softening turned off. 
With gravitational softening, dynamical interactions between the stars are not captured properly, and the velocities of the stars will be lower as a result. 
We performed an additional no softening model but with grouping parameters the same as the `standard' model (not shown here) to confirm that the mass distribution also plays a role in determining the velocity dispersion as expected.

\begin{figure}
    \centering
    \includegraphics[width=\columnwidth, trim={0 2cm 0 1cm}]{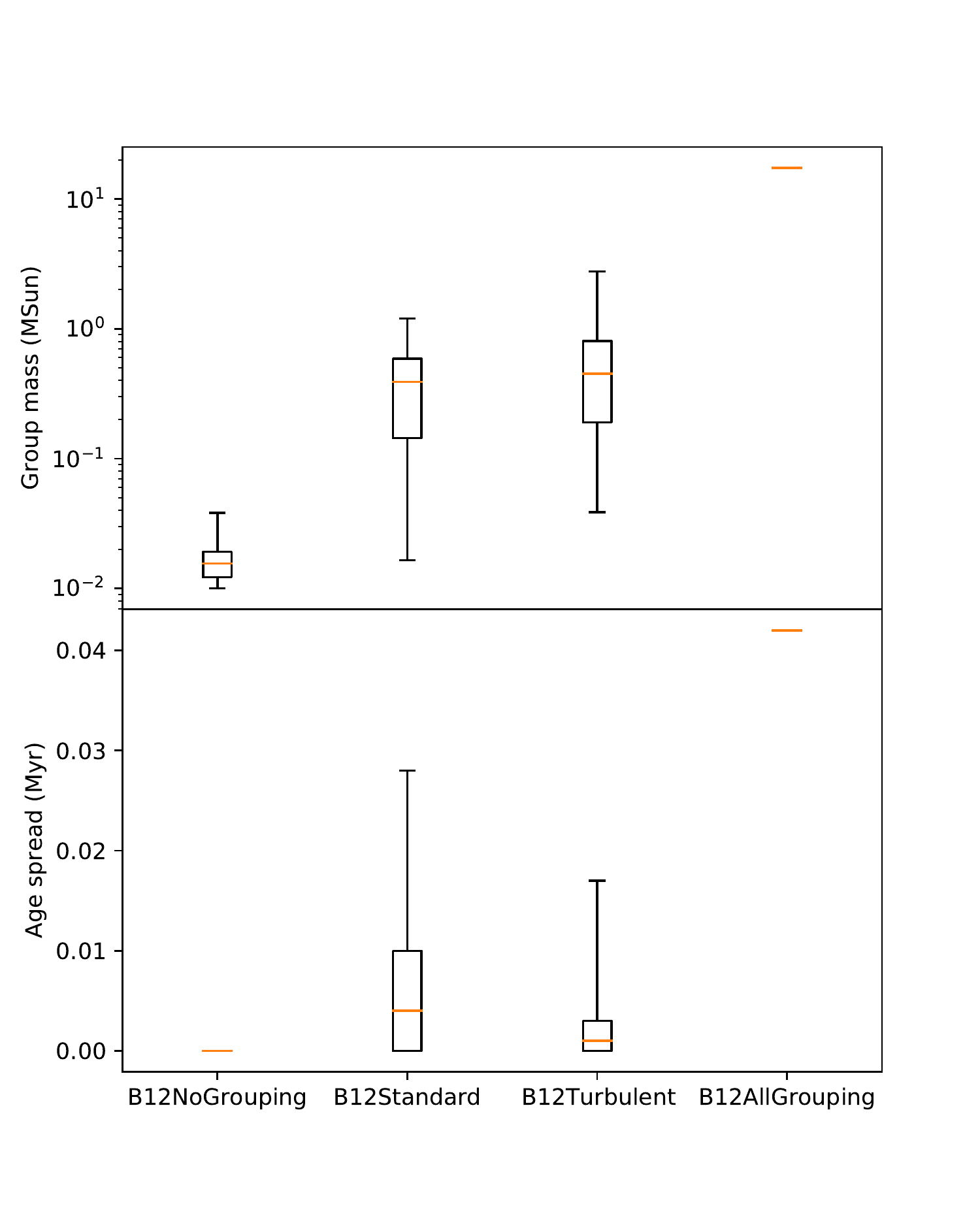}
    \caption{The group mass and age spread of the star groups for each model at the free-fall time of 0.19 Myr are shown. 
    For each box, the orange line is the median group mass, the length of the box is the interquartile range, and the whiskers span the whole range.}
    \label{fig:B12GA}
\end{figure}

Figure \ref{fig:B12GA} shows the mass and age spread of the star groups (recall that sinks are first grouped together, then stars form in each group) for the `no grouping', `standard', `turbulent', and `all grouping' models at the free-fall time of 0.19 Myr. 
The models are arranged in increasing degree of grouping, which can be roughly deduced from the star mass fraction $f$ shown in Table \ref{tab:initialconditions}. 
In general, as the degree of grouping increases, we see an increase in the median mass of star groups. 
For the `no grouping' model, a great proportion of mass is `trapped' in the sinks and is unable to form stars, which causes the low star group masses. 
We do not see a clear trend in the group age spread in our models in comparison of this length and time scales. 
There is no age spread for the groups in the `no grouping' model as they are single-member groups. 
On the other hand, for the `all grouping' model, the age spread of the star group simply shows the duration between the first star formation and the current free-fall time. 
We also analyse the virial radius and the velocity dispersion of the star groups, and they simply reflect the length scale ($\sim 10^{-2}$ pc as shown in Figure \ref{fig:B12SD}) and speed scale ($\approx 3$ km/s, same as the turbulence of the cloud) of the star groups.

\subsection{Parsec-scale cloud-cloud collision simulation}
\label{ssec:resultl20}

\begin{figure}
    \centering
    \includegraphics[width=\columnwidth]{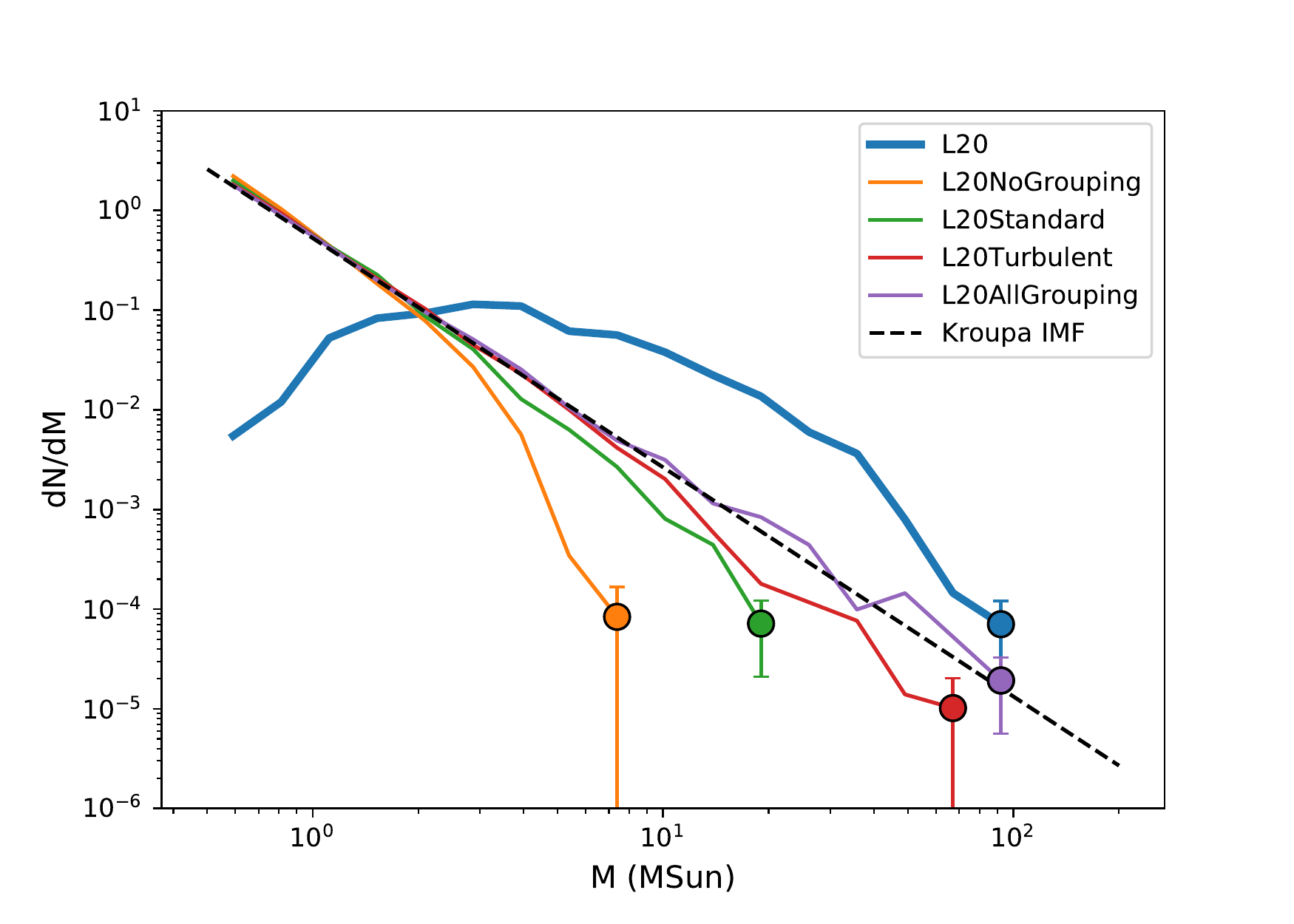}
    \caption{The IMF of our models are compared with that of \protect\cite{liow_collision_2020}, labelled as L20, at 1.75 Myr, the time when 10\% of the gas mass is converted to sink mass in the original simulation. 
    The markers are the maximum mass bin in the respective models. 
    The errorbars are the standard deviations from Poisson sampling. 
    The black dashed line is the analytical Kroupa IMF.}
    \label{fig:L20IMF}
\end{figure}

Figure \ref{fig:L20IMF} shows the IMFs of all the models that compare with the chosen simulation from \cite{liow_collision_2020}, labelled L20 (blue) at 1.75 Myr, the time when 10\% of the gas mass is converted to sinks and stars, the same timescale shown in the figures in \cite{liow_collision_2020}. 
Although the L20 simulation has a much lower resolution than B12 (see Section \ref{ssec:resultb12}), the form of the sink mass function is not dissimilar to the Kroupa IMF at high masses; at low masses the mass function departs strongly from the Kroupa IMF.
Firstly, the `no grouping' model (orange) that uses \textit{single} star formation is not able to form any high mass stars > 10 \msun{}, and deviates significantly from the Kroupa IMF at the high mass range. 
With \textit{grouped} star formation, the resultant IMFs are closer to the Kroupa IMF, however the `standard' model (green) with $d\_g=1$ pc and $v\_g=1$ km/s is still not enough to sample a complete Kroupa IMF. 
Similar to the comparison with \cite{bate_cluster_2012}, both the `turbulent' model (red) that uses the initial turbulent velocity as the grouping speed, and the `all grouping' model (purple) can sample the full Kroupa IMF. 
However the `turbulent' model does not need to group all the sinks together like the `all grouping' model, which as we shall see is beneficial for larger scale simulations. 
The star mass fraction $f=0.901$ of the `turbulent' model is also high compared to other models. 


\begin{figure*}
    \centering
    \includegraphics[width=\textwidth]{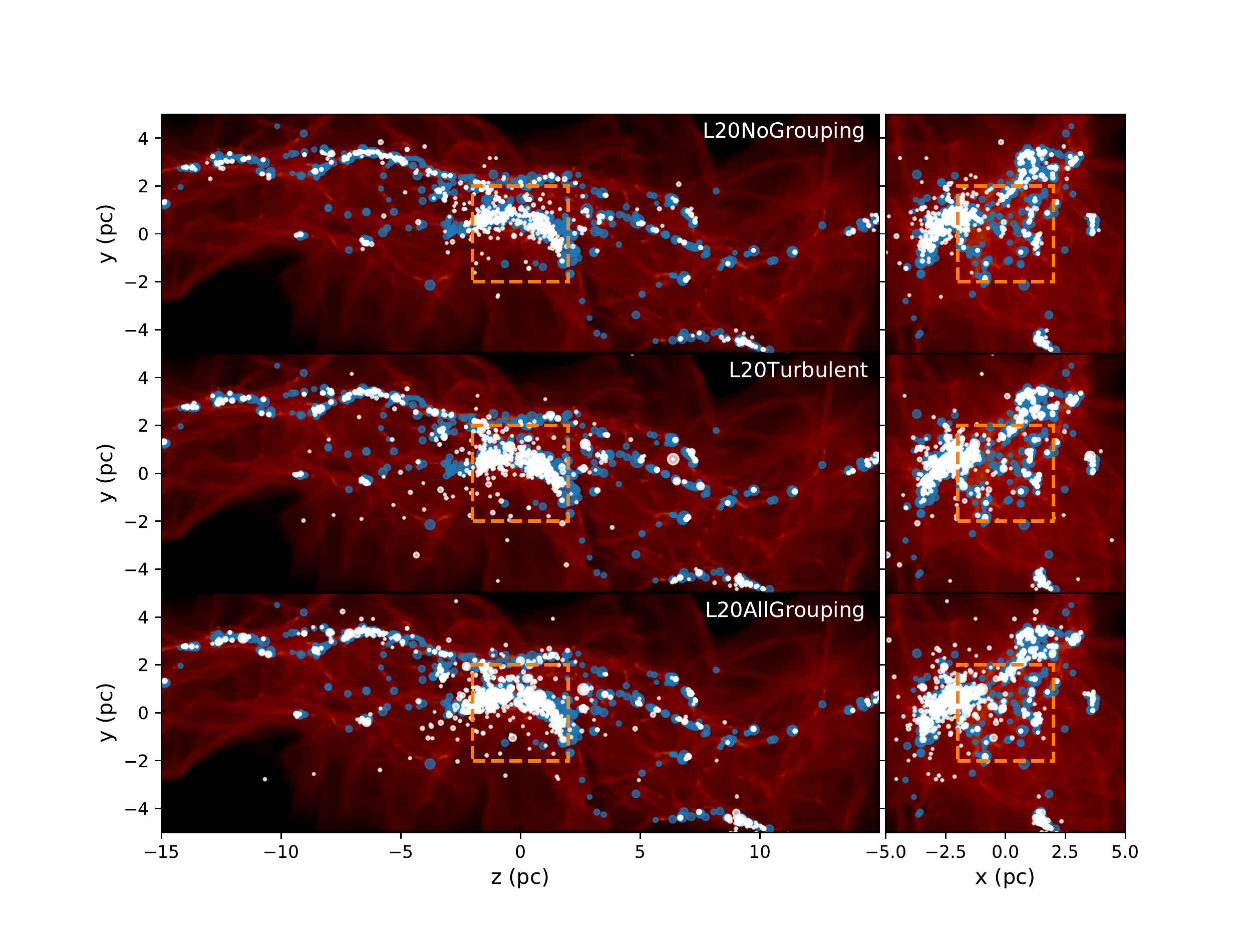}
    \caption{The stars from the `no grouping', `turbulent', and `all grouping' models (white) are compared against the sinks from \protect\cite{liow_collision_2020} (blue) at the 1.75 Myr. 
    The marker radius is proportional to the particle mass. 
    The gas distribution from the original simulation is set as background for reference, as the gas distribution from the individual models are visually identical to the original one.
    The orange dashed box shows the region of interest for the discussion of local mass conservation.}
    \label{fig:L20SD}
\end{figure*}

Figure \ref{fig:L20SD} shows the stellar distribution of the `no grouping', `turbulent', and `all grouping' models plotted over the sink distribution of L20. 
Since each sink in L20 is best approximated as a small group of stars, the size of the sinks are exaggerated as compared to the stars from our models. 
For the `no grouping' model (top plot), the stars are located about the dense gas filaments, however their masses are small as noted in Figure \ref{fig:L20IMF}. 
As the degree of grouping in \textit{grouped} star formation increases, more higher mass stars that are comparable to the original simulations are formed, again already inferred in Figure \ref{fig:L20IMF}. 
However, at the same time, we note that the stellar distribution reflects the true dense gas filament distribution less as grouping increases. 
This means that local mass conservation, i.e. within scales of order parsecs, becomes more violated as the degree of grouping increases.
In the case of the `all grouping' model (bottom plot) where all sinks are grouped as one, mass is unphysically transferred to the clustered area from less clustered regions. 
Similar unphysical transfer of mass to some extent can occur using the star formation schemes described in \cite{fujii_initial_2015} and \cite{smith_sensitivity_2021}. 
This may be unwanted if the length resolution of a simulation is parsec or sub-parsec, whereas in the comparison with \cite{bate_cluster_2012} in Section \ref{ssec:resultb12}, the size scale of the whole system is still small compared to, for example, whole galaxy simulations. 

To quantify mass conservation, we take a box of length 4 pc centred at the origin, and calculate the percentage of total sink and star mass from each model ($M\_{sinks,model,box} + M\_{stars,model,box}$) on the total sink mass from L20 ($M\_{sinks,L20,box}$). 
The location of the box contains about half of the most massive cluster obtained from the original simulation of L20 (see \cite{liow_collision_2020} for more information on the identification of this massive cluster, labelled `L4'), ideal to investigate local mass conservation. 
We find that for the `no grouping' model, 97\% of the mass is conserved locally, however only about 65\% of the locally conserved mass are stars as this model does not form many massive stars, and a lot of mass is still `trapped' in the sinks without any star formation. 
On the other hand, even though the `all grouping' model is extremely efficient in converting sink mass to star mass ($f=0.999$, so $M\_{sinks}$ and in extension $M\_{sinks,model,box}$ are practically negligible in this model) and producing massive stars, only about 40\% of mass is conserved locally. 
For the `turbulent' model (middle plot), about 90\% of the mass is conserved locally, and about 90\% of the locally conserved mass is contributed by stars, suggesting that this model is the most successful in terms of forming a similar mass of stars to the original model, reproducing the Kroupa IMF, and obeying local mass conservation.

\begin{figure}
    \centering
    \includegraphics[width=\columnwidth]{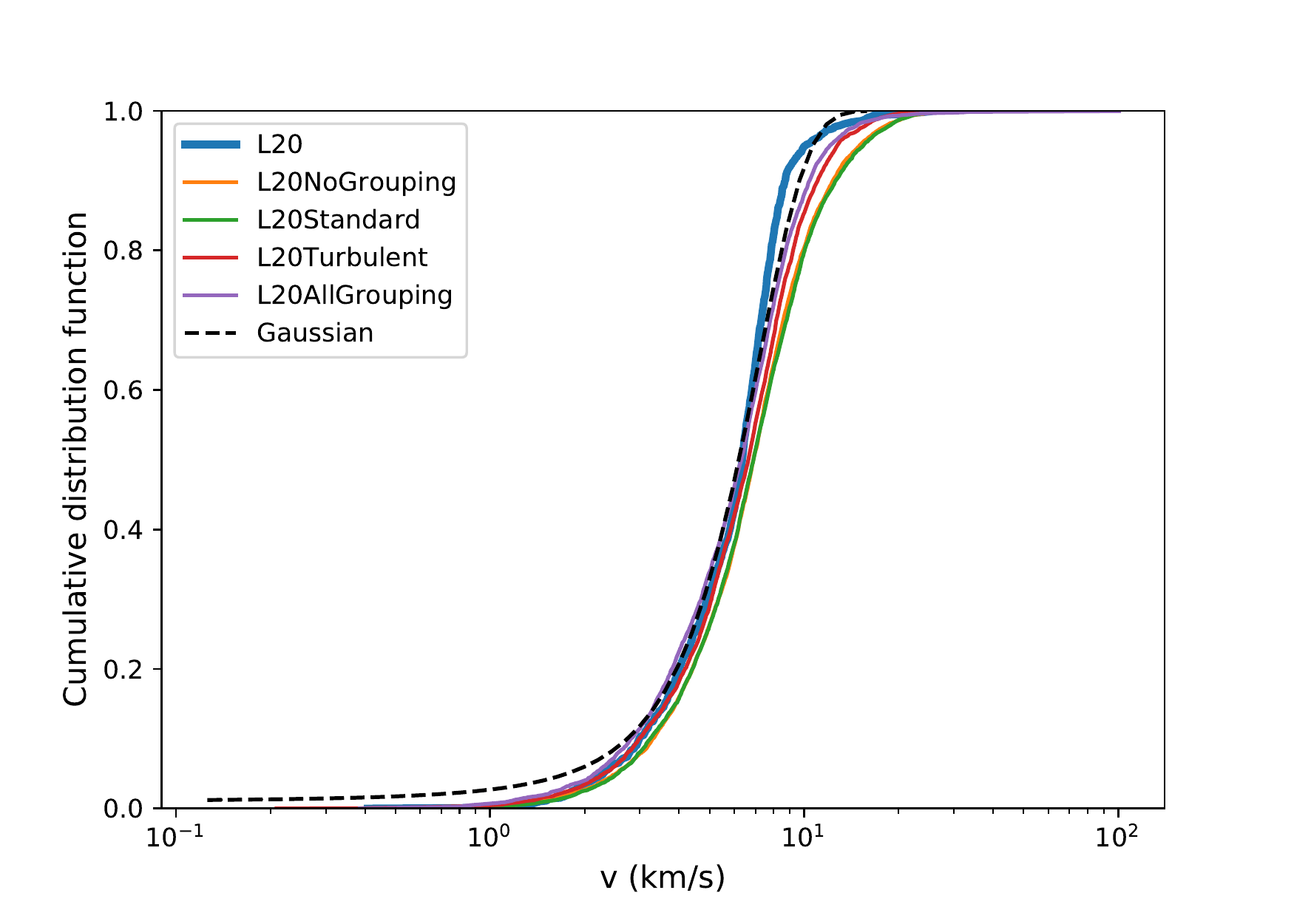}
    \caption{The cumulative velocity distribution of the models are shown are compared with that of \protect\cite{liow_collision_2020} at 1.75 Myr. 
    The black dashed line is the Gaussian cumulative distribution function, fitted on the velocity distribution of the original simulation, with mean 6.2 km/s and standard deviation of 2.7 km/s.}
    \label{fig:L20VD}
\end{figure}

Figure \ref{fig:L20VD} shows the cumulative velocity distribution function for all models compared to that of L20. 
The velocity dispersion of L20 is fitted with a Gaussian distribution with mean 6.2 km/s and standard deviation of 2.7 km/s (black dashed line). 
As large sample sizes cause bias towards rejecting the null hypothesis for equal distributions, KS tests are not suitable for hypothesis testing here \citep{phacking_large_sample_2021}. 
Nonetheless, all velocity distributions are visually Gaussian-like and similar to that of L20, regardless of the grouping parameters, suggesting that the velocity distribution of the larger scale models is relatively independent of the details of the star formation prescription. 


Similar to Figure \ref{fig:B12GA}, we analyse the properties of the star groups in this comparison, however in general we find similar conclusions to those of Section \ref{ssec:resultb12}. 
The mass of the star groups increases with the degree of grouping (which is deduced from the star mass fraction $f$) increases. 
Similarly, the group age spread shows no clear correlation with the models. 
The virial radius of the groups ranges from $\sim 10^{-1} - 10^0$ pc, reflecting the length scales of star-forming regions seen in Figure \ref{fig:L20SD}, while the velocity dispersion of the star groups is $\approx 6$ km/s, 
arising from the collision and initial turbulence. 

\subsection{Parsec-scale isolated cluster simulation}
\label{ssec:resultsj}


\begin{figure}
    \centering
    \includegraphics[width=\columnwidth]{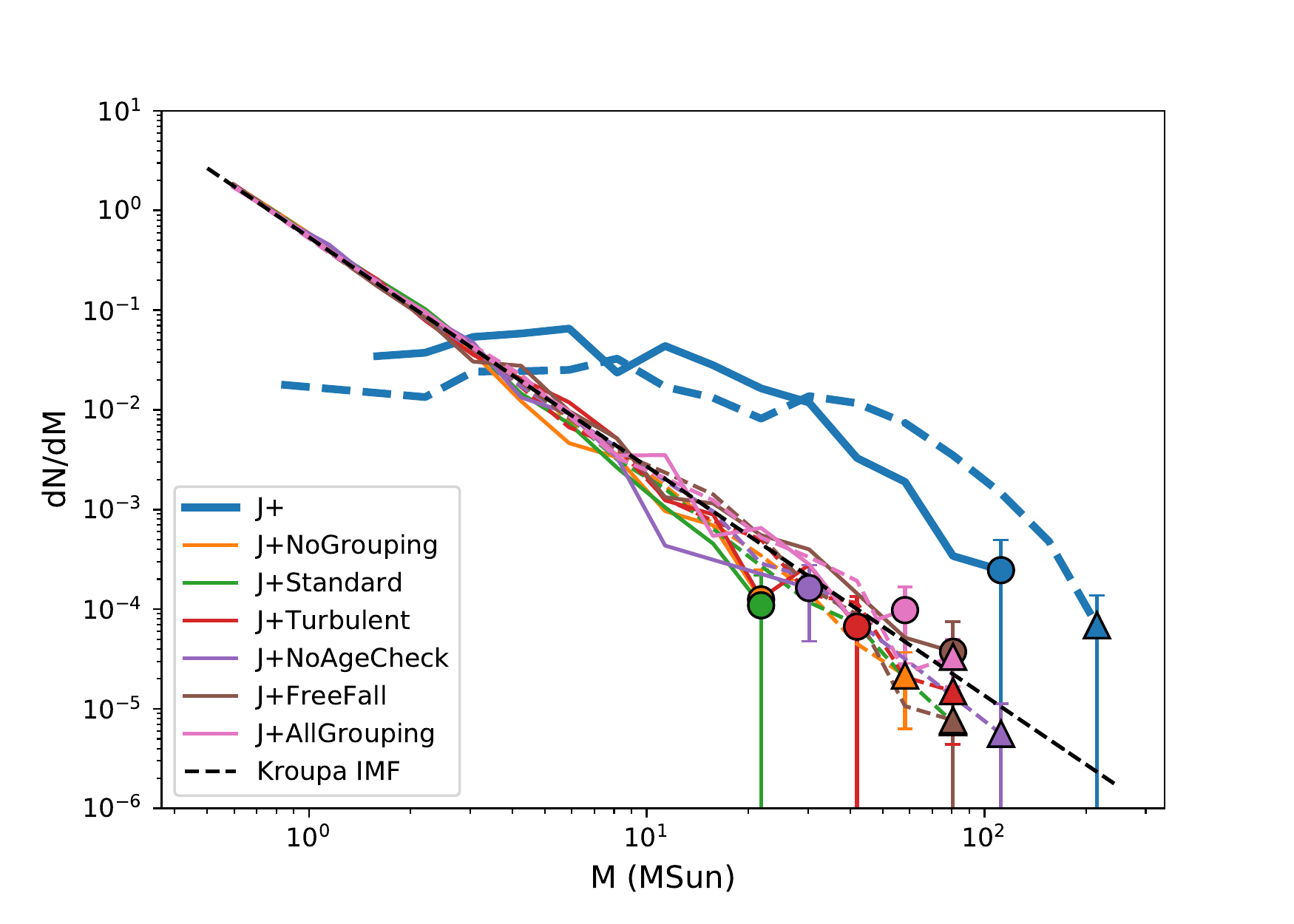}
    \caption{The IMF of our models are compared with that of the original simulation by \protect\cite{jaffa_sims_prep}, labelled as J+, at 5 Myr (solid lines with circle ends) and at 20 Myr (dashed lines with triangle ends). 
    The markers are the maximum mass bin in the respective models. 
    The errorbars are the standard deviations from Poisson sampling. The black dashed line is the analytical Kroupa IMF. 
    } 
    \label{fig:SJIMF}
\end{figure}

Figure \ref{fig:SJIMF} shows the IMF of the different models as compared with the original simulation by \cite{jaffa_sims_prep}, labelled J+ at 5 Myr and at 20 Myr.
In this comparison, we do not rerun the original simulation for practical reasons explained in Section \ref{sssec:icjaffa}. 
If we were to rerun the simulation, stars would form dynamically and a fraction of sink mass is constantly converted to stars, leaving the sinks smaller than the sinks in J+ at any time. This means that we would expect more low mass stars than the IMFs shown in Figure \ref{fig:SJIMF} if stars are formed dynamically.
Nonetheless, at 5 Myr, the maximum star mass achievable by the models with less to no grouping, e.g. the `no grouping' (orange) and `standard' (green) models are generally smaller than those with greater degree of grouping, the extreme case being the `all grouping' model (pink). 
Once again, this is simply because greater grouping allows for more massive mass reservoir to form higher mass stars. 
Similar to our inference from Section \ref{ssec:resultb12}, the `turbulent' model (red), i.e. setting $d\_g = 1$ pc, $v\_g=3$ km/s (about the turbulence speed), and $\tau\_g = t\_{ff} = 5.24$ Myr is able to form high mass stars.
At 20 Myr, the maximum star mass achievable by all models converges towards 100 \msun{}, the upper star mass limit of the Kroupa IMF. 
For the solid lines in Figure \ref{fig:SJIMF}, 5 Myr $< t\_{ff} = 5.24$ Myr, so the age criterion is not used at this point.

\begin{figure}
    \centering
    \includegraphics[width=\columnwidth, trim={0 2cm 0 1cm}]{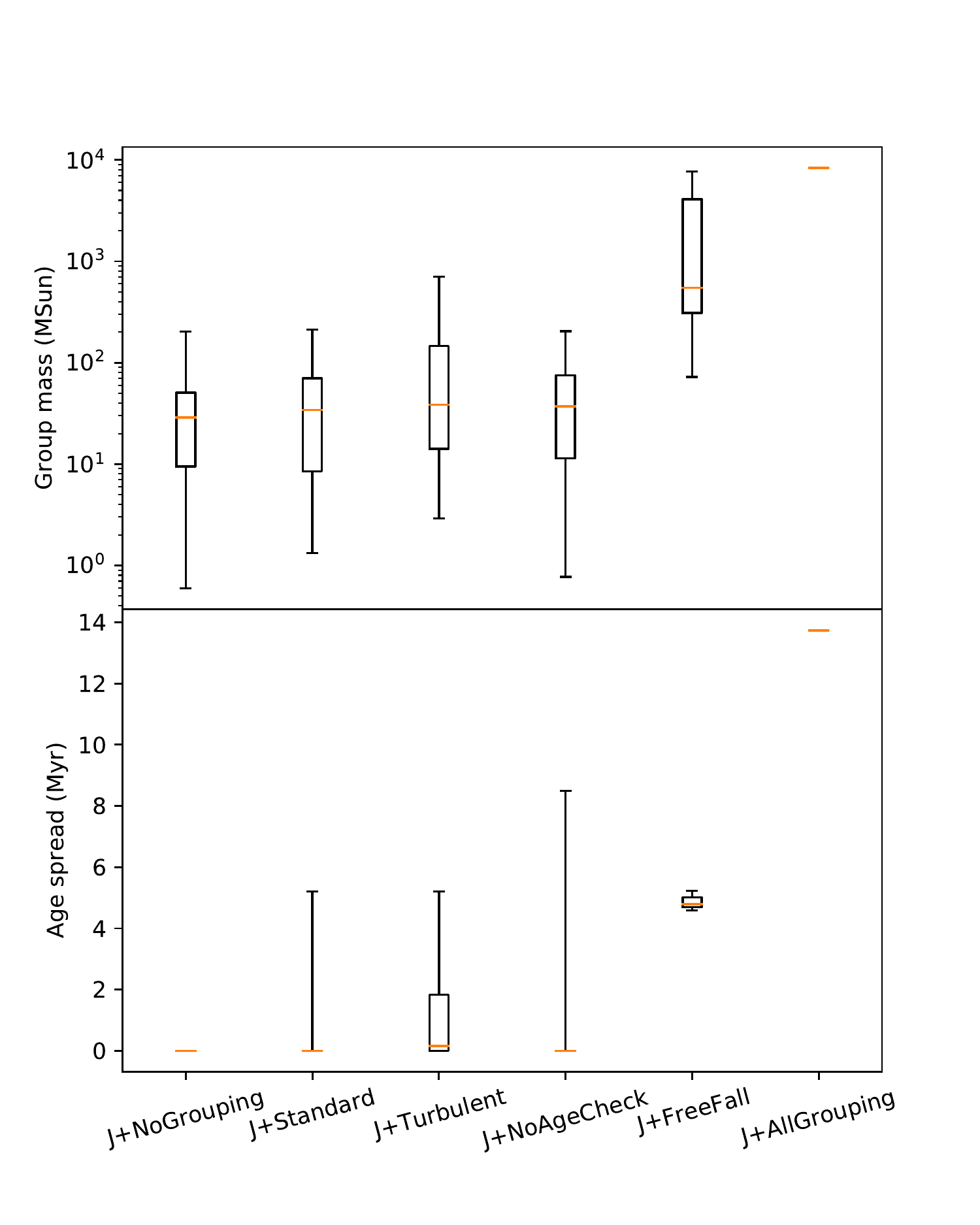}
    \caption{The mass and age spread of the groups for each model at 20 Myr. 
    For each box, the orange line is the median group mass, the length of the box is the interquartile range, while the whiskers span the whole mass range. 
    }
    \label{fig:SJGA}
\end{figure}

Figure \ref{fig:SJGA} shows the mass and age spreads of the groups at 20 Myr, where the age criteria do become relevant. 
In the top subplot, even though the median star group mass of the `turbulent' model is comparable to other models that do not group all sinks, the upper limit for the star group mass is greater, allowing the formation of larger mass stars due to the larger mass reservoir. 
The `no age check' model, which differs from the `standard' model in its grouping age parameter $\tau\_g$, shows a similar number of groups (Table \ref{tab:initialconditions}) and group mass distribution to the `standard' model. 
This suggests either the grouping age criteria generally do not affect the degree of grouping, or the system needs to evolve much longer to allow the age criteria to take effect. 
The `free-fall' model, which differs just with the velocity criterion, shows a greater median mass than the `no age check' and `standard' models, showing that the grouping distance and speed parameters affect the grouping of sinks more significantly than the grouping age parameter.

The bottom subplot of Figure \ref{fig:SJGA} shows the age spread of the sink groups prior to star formation at 20 Myr. 
Our modified star formation scheme in this comparison does not form stars dynamically, so all stars form at 20 Myr with zero age spread.
Similar to Figure \ref{fig:B12GA}, there is no age spread for the sink groups in the `no grouping' model simply because each group consists of only one sink by construction, and also the age spread of the sink group in `all grouping' model is simply the time between first sink formation and the current snapshot of 20 Myr. 
For the `standard' model, only 32 out of 160 (20\%) sink groups have non-zero age spreads, and among them only 3 groups have age spread $>t\_{ff}/2 = 2.62$ Myr. 
The age spread distribution of the sink groups in the `turbulent' model is more skewed towards greater values. 
In both models, the maximum age spread is about $t\_{ff}=5.24$ Myr, the value of the grouping age parameter $\tau\_g$. 
For the `no age check' model, 36 out of 155 (23\%) sink groups have non-zero age spreads, similar to the `standard' model. 
However, because the grouping age criteria is neglected, the maximum age spread can go beyond $t\_{ff}$, and in this case 5 sink groups have age spread beyond $t\_{ff}$, the maximum value being 8.49 $\approx 1.6 t\_{ff}$.
This is undesirable because each small star-forming region (i.e. group) is expected to form stars within one free-fall timescale, i.e. the formation timescale under gravitational collapse. 
Lastly, the `free-fall' model creates 3 groups at 20 Myr, all of which have age spread around $t\_{ff} = 5.24$ Myr as expected since sinks can be grouped freely in position and velocity but restricted temporally, so this model can create at most 3 groups $(\lfloor 20/5.24 \rfloor = 3)$. 
In conclusion, even though the grouping age criteria does not seem to alter the group masses much, it is useful to ensure consistent age spreads among the groups. 
This criteria is essential in star-forming simulations that run much longer than the free-fall or any formation timescales. 

\subsection{Kiloparsec-scale spiral arm simulation}
\label{ssec:resultrieder}

\begin{figure}
    \centering
    \includegraphics[width=\columnwidth]{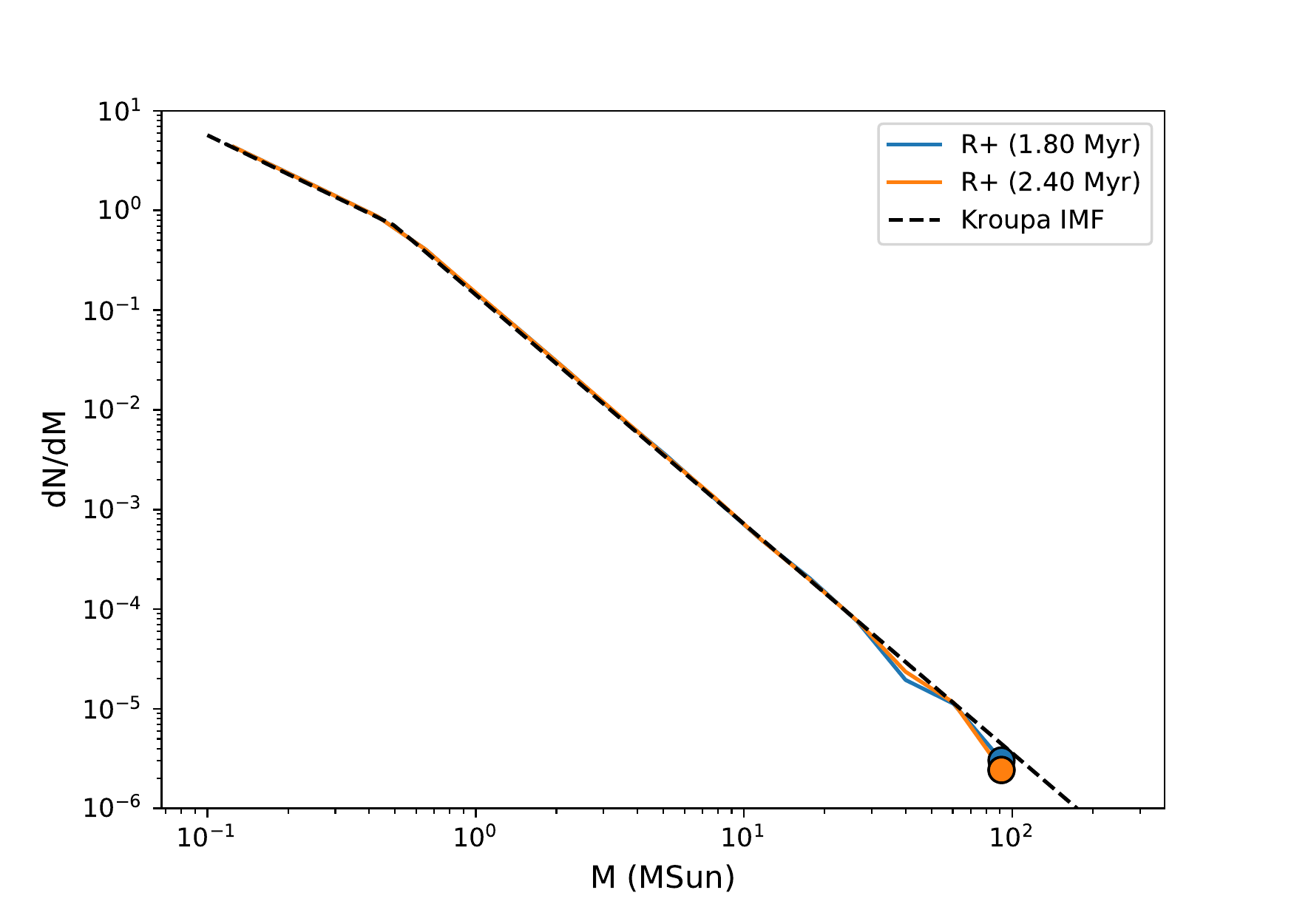}
    \caption{The IMF of the sub-galactic simulation by \protect\cite{rieder_ekster_2021} at 1.80 Myr and 2.40 Myr are shown. 
    The black dashed line is Kroupa IMF. 
    The errorbars are the standard deviations from Poisson sampling.
    The black dashed line is the analytical Kroupa IMF.}
    \label{fig:SRIMF}
\end{figure}

Figure \ref{fig:SRIMF} shows the IMF of the stars from a simulation by \cite{rieder_ekster_2021} at 1.80 Myr and at 2.40 Myr. 
Even though the stars are formed using the \textit{single} star formation method, i.e. no group is assigned, the IMFs are already close to a complete Kroupa IMF. 
In this low mass resolution simulation, the mass reservoir for star formation, i.e. the average sink mass in this case, is $\sim 10^2$ \msun{}, enough to sample a complete IMF. 
This confirms that if the average sink mass is large, then we can fully sample the IMF, which would tend to be the case for larger galaxy or even cosmological simulations. 
In these simulations, the \textit{single} star formation method is appropriate.

\section{Discussion}
\label{sec:discussion}

\subsection{
Effect of changing the random seed on cluster properties}
\label{ssec:randomness}

\begin{figure}
    \centering
    \includegraphics[width=\columnwidth]{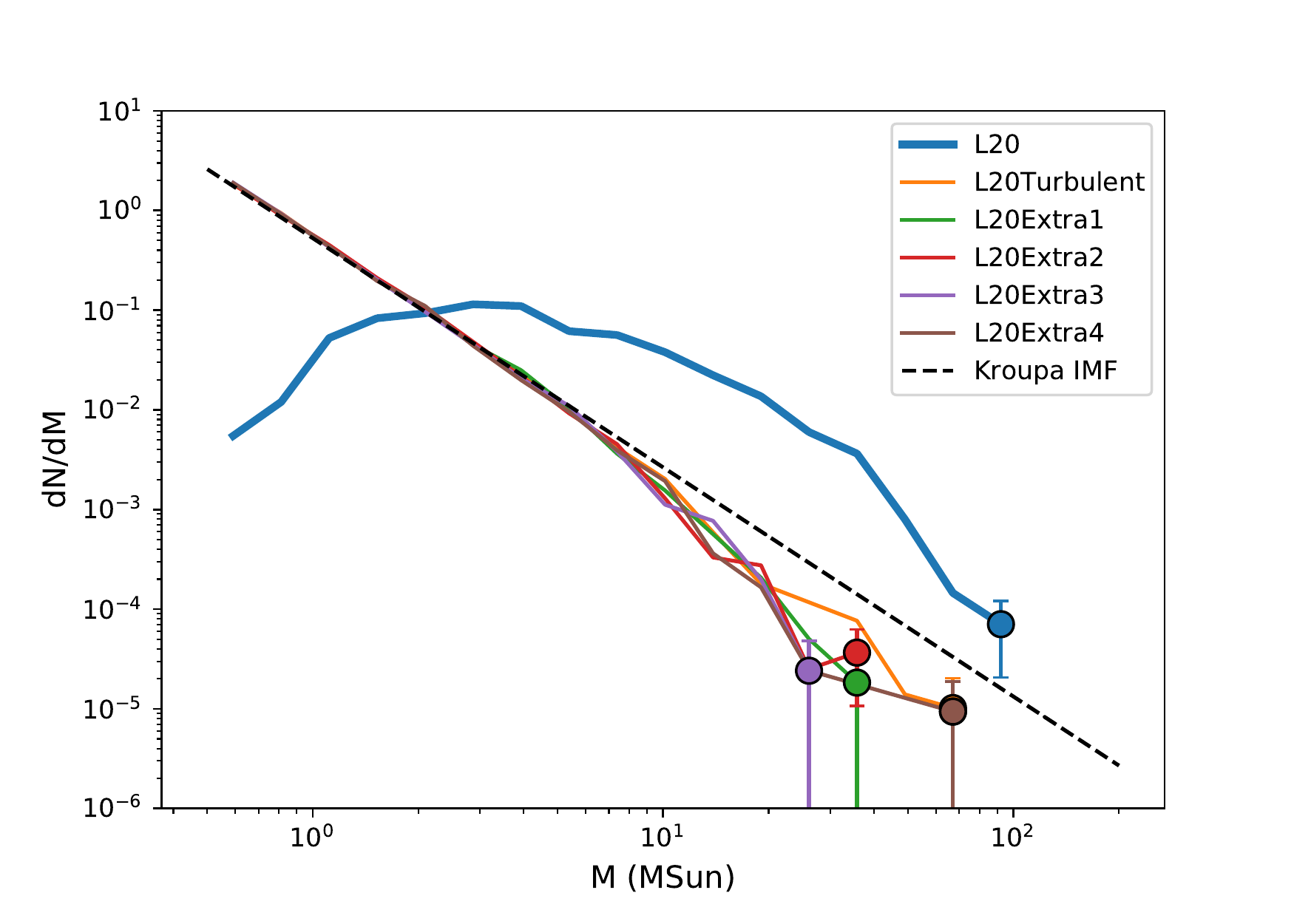}
    \caption{The IMF of the reruns with different random seeds for star formation are compared with that of \protect\cite{liow_collision_2020}, labelled as L20, and the `turbulent' model in Section \ref{ssec:resultl20}, labelled as L20Turbulent, at 1.75 Myr, the time when 10\% of the gas mass is converted to sink mass in the original simulation. The markers are the maximum mass bin in the respective models. The errorbars are the standard deviations from Poisson sampling. The black dashed line is the analytical Kroupa IMF.}
    \label{fig:extraIMF}
\end{figure}

In our simulations, the stars' initial positions are allocated within the accretion radius and the velocities according to the gas dispersions, but we do use a random seed to assign individual positions and velocities. Here we determine whether properties of the clusters formed have any dependence on the random positions and velocities.
To test the effect of randomness on our results, we choose the L20Turbulent model  (Section \ref{ssec:resultl20}) and perform four more simulations with the exact same initial conditions, except we set a different random seed for star formation compared to the original `turbulent' model. This leads to different positions of the stars, and a different velocity field.
Figure \ref{fig:extraIMF} shows the IMF of the `turbulent' colliding clouds model and the reruns at 1.75 Myr. 
Even though there are slight deviations between the IMFs, changing the random seeds for star formation does not alter the overall trend of the mass function. 
Other results, such as the cumulative velocity distribution and the properties of the groups of the reruns are also similar compared to those of the `turbulent' model. Using the KS tests as described in Section \ref{ssec:resultl20}, we find that the velocity distributions may be considered statistically the same.


\begin{figure}
    \centering
    \includegraphics[width=\columnwidth]{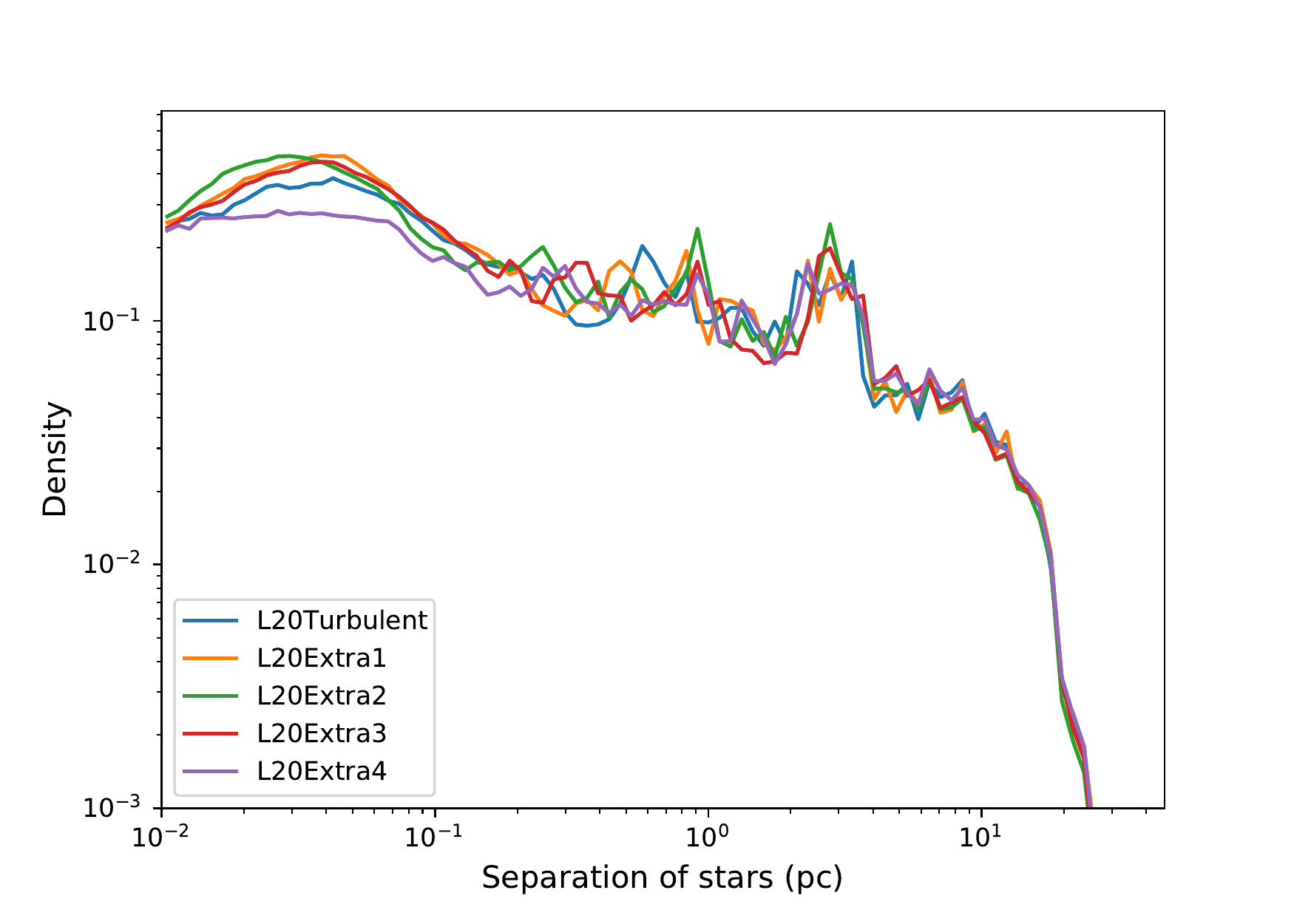}
    \caption{The density histograms of the separation of stars for the reruns are compared with that of the `turbulent' model in Section \ref{ssec:resultl20} at 1.75 Myr.}
    \label{fig:separation_extra}
\end{figure}

\begin{figure*}
    \centering
    \includegraphics[width=\textwidth]{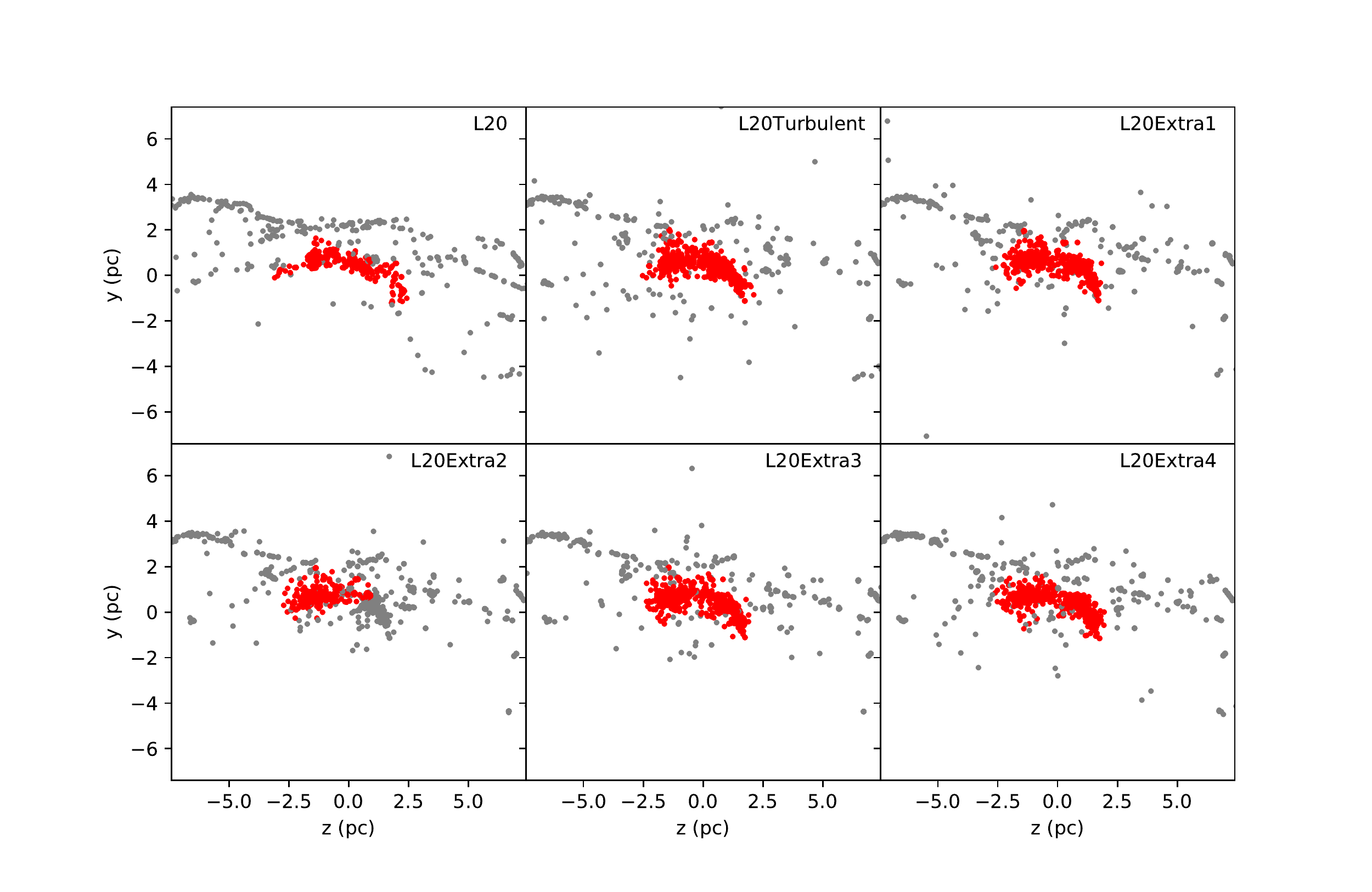}
    \caption{The stellar distribution of the simulation by \protect\cite{liow_collision_2020}, labelled as L20, the `turbulent' model in Section \ref{ssec:resultl20}, labelled as L20Turbulent, and the reruns at 1.75 Myr are shown with the most massive clusters (red) identified by DBSCAN.}
    \label{fig:cluster_extra}
\end{figure*}

The stellar distributions of the reruns are visually indistinguishable from that of the `turbulent' model (middle plot in Figure \ref{fig:L20SD}), but we perform additional quantitative analysis to show the similarity between the stellar distributions. 
Figure \ref{fig:separation_extra} shows the density histograms of the separation of stars for the reruns and the `turbulent' model at 1.75 Myr, similar to the analysis by \cite{torniamenti_clustering_2021} to study the spatial distribution of the stars. 
Subtle differences can be observed among the five models, but generally they share similar overall trends and features such as the peaks at about 1 pc and 3 pc, signifying similar length scales of clustering. 
We use the same density-based clustering algorithm \citep[DBSCAN; ][]{ester_density_based_1996} and parameters (i.e. maximum separation between members of 0.5 pc and minimum number of members of 5) used in \cite{liow_collision_2020} to identify the most massive cluster in each of the `turbulent' model and the reruns. 
These clusters are shown in Figure \ref{fig:cluster_extra} which shows that by eye the distribution of stars and grouping into clusters is very similar. 
Except for the second rerun (labelled as L20Extra2), the clustering algorithm identifies the same most massive cluster in the reruns and in the `turbulent' model.  
The most massive cluster in the second rerun is smaller simply because its stellar distribution is slightly more fragmented compared to the other models by chance. 
By increasing the maximum separation between members slightly to 0.6 pc, the most massive cluster identified in the second rerun becomes the same as the rest, showing that these clusters are statistically indifferent.
Therefore, we find that cluster formation and evolution are relatively independent of the stochasticity introduced in our star formation prescription.

\subsection{Varying the upper star mass limit}
\label{ssec:varying_limit}

So far in this paper, we fix a constant value of 100 \msun{} as the upper star mass limit 
to our input Kroupa IMF, which is used to sample the stellar population and decide the mass threshold to form further stars.
In the case where the group masses are sufficiently large, most of the mass in the sinks can be converted to stars, for example for our comparisons in Sections \ref{ssec:resultl20}, \ref{ssec:resultsj} and \ref{ssec:resultrieder}. 
However in Section \ref{ssec:resultb12}, the star mass fraction can be very low, which is clearly not ideal. 
If we take the extreme case, the `no grouping' model, the average group mass (each group is equivalent to a sink particle in this case) is about 0.02 \msun{} $<$ 0.37 \msun{}, the expected mass value calculated using the input IMF and mass range. Probabilistically, even relatively low mass stars cannot be assigned to the sinks, and consequently much of the stellar mass remains `trapped' in the sinks rather than being converted to stars.

\begin{table}
    \centering
    \begin{tabular}{l|l|l}
    \hline \hline
    Model            & $f_1$ & $f_2$ \\
    \hline \hline 
    B12NoGrouping    & 0.050 & 0.739 \\
    B12Standard      & 0.498 & 0.864 \\
    B12Turbulent     & 0.576 & 0.935 \\
    B12AllGrouping   & 0.766 & 0.995 \\
    B12NoSoftening   & 0.597 & 0.983 \\
    B12LowResolution & 0.987 & 0.988 \\
    \hline 
    L20NoGrouping    & 0.708 & 0.763 \\
    L20Standard      & 0.804 & 0.894 \\
    L20Turbulent     & 0.901 & 0.974 \\
    L20AllGrouping   & 0.999 & 0.999 \\
    \hline \hline
    \end{tabular}
    \caption{The star mass fraction of the models when the upper star mass limit is equal to 100 \msun{} $f_1$ and the individual group mass $f_2$. The column $f_1$ is identical to the column $f$ in Table \ref{tab:initialconditions}.}
    \label{tab:group_equal_limit}
\end{table}

An alternative is to set the upper star mass limit equal to the individual group mass.
Besides eliminating a parameter from our star formation prescription, this new setting ensures that the mass threshold and the star mass that can form within the group are always less than the group mass, avoiding the problem of having the mass threshold and stellar masses larger than the mass available locally. 
Consequently, the sink-to-star conversion is more efficient as shown in Table \ref{tab:group_equal_limit}, where we test this setting by rerunning the models that compare with B12 (Section \ref{ssec:resultb12}) and L20 (Section \ref{ssec:resultl20}). 
The star mass fraction of the reruns in B12,
especially those with lower degree of grouping, is boosted using the new setting. 
Using the group mass as the upper limit of the IMF has less impact on the star mass fraction of the reruns of L20, as the sink-to-star conversion is already relatively efficient in those models.

\begin{figure}
    \centering
    \includegraphics[width=\columnwidth]{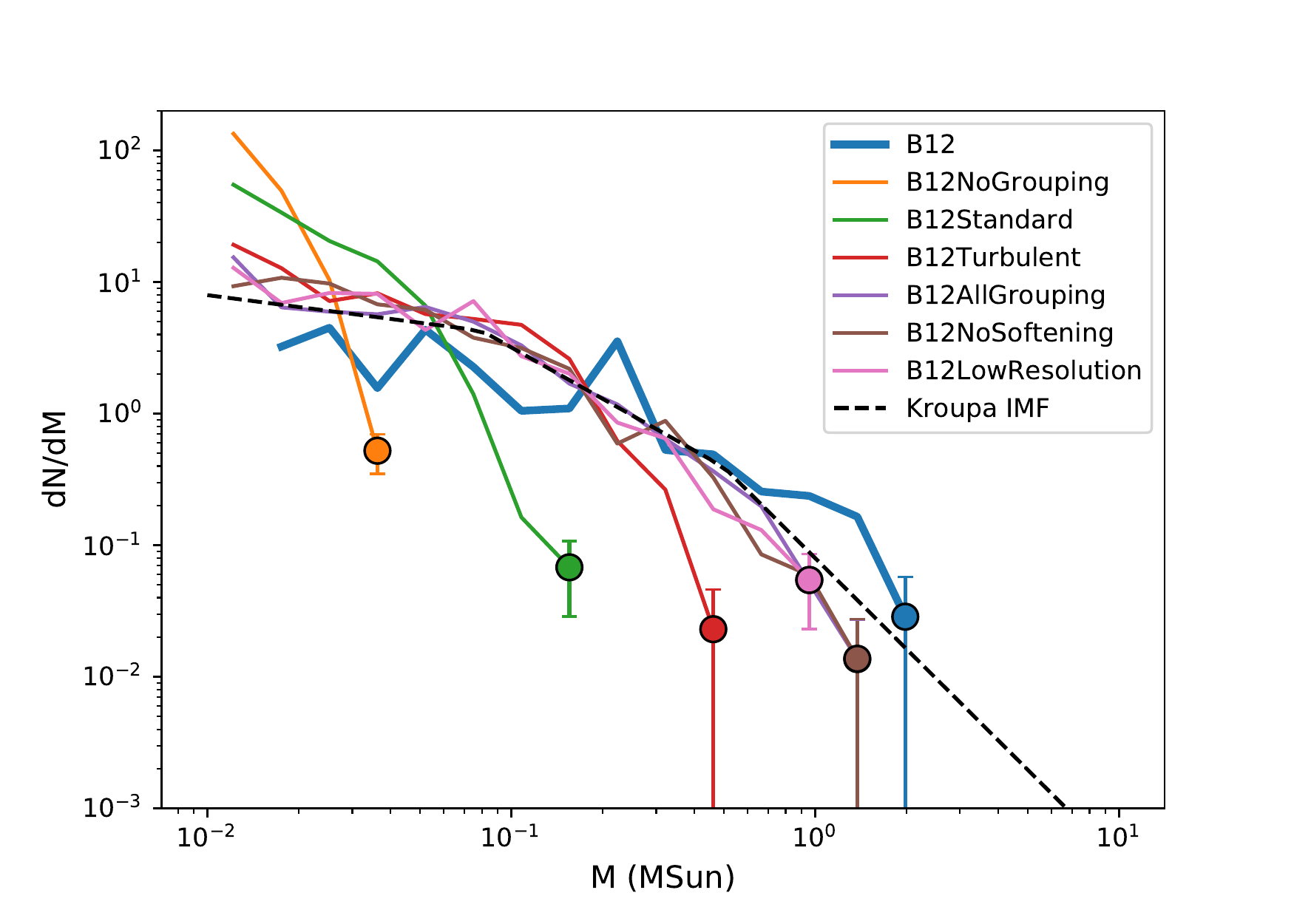}
    \caption{The IMF of the reruns when the upper star mass limit equals to individual group mass are compared with that of \protect\cite{bate_cluster_2012}, labelled as B12, at the free-fall time of 0.19 Myr. 
    The markers are the maximum mass bin in the respective models.
    The errorbars are the standard deviations from Poisson sampling.
    The black dashed line is the analytical Kroupa IMF.}
    \label{fig:B12IMF_group_equal_limit}
\end{figure}

\begin{figure}
    \centering
    \includegraphics[width=\columnwidth]{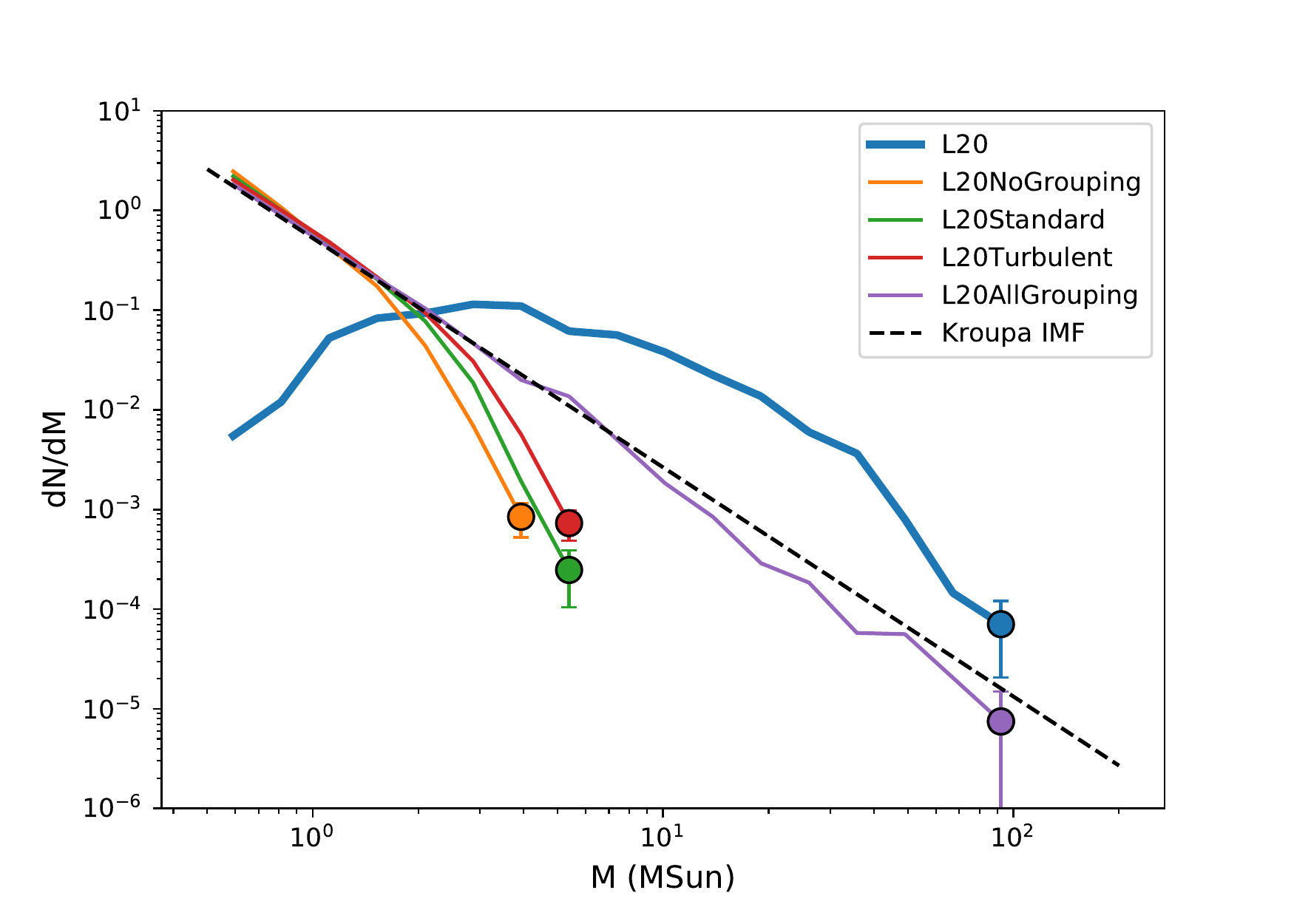}
    \caption{The IMF of the reruns when the upper star mass limit equals to individual group mass are compared with that of \protect\cite{liow_collision_2020}, labelled as L20, at 1.75 Myr, the time when 10\% of the gas mass is converted to sink mass in the original simulation. 
    The markers are the maximum mass bin in the respective models. 
    The errorbars are the standard deviations from Poisson sampling. 
    The black dashed line is the analytical Kroupa IMF.}
    \label{fig:L20IMF_group_equal_limit}
\end{figure}

One major disadvantage of using the group mass as the upper mass limit is that more low mass stars, and subsequently less high mass stars, are created since the small individual upper star mass limits prevent the formation of higher mass stars. 
Figures \ref{fig:B12IMF_group_equal_limit} and \ref{fig:L20IMF_group_equal_limit} show the IMF of the reruns as compared to B12 and L20 using the group mass as the upper star mass limit. 
In a smaller scale system as shown in Figure \ref{fig:B12IMF_group_equal_limit}, all the reruns have more low mass stars as compared to the models in Figure \ref{fig:B12IMF}. The most massive stars in the reruns are also less massive compared to their respective counterparts. 
In the larger scale L20 system as shown in Figure \ref{fig:L20IMF_group_equal_limit}, all models except the `all grouping' case cannot form massive stars $>$ 5 \msun{} (c.f. Figure \ref{fig:L20IMF}). 
In the previous models that used 100 \msun{} as the upper star mass limit, the group masses in these large scale models are generally more massive, but note that these sink groups are usually small upon creation. 
Therefore, taking the group mass as upper star mass limit instead, these sink groups would form stars right after sink creation and become less massive.
Since the density of the sinks is kept constant, their accretion radii become shorter and therefore prevent the groups to grow in mass and form higher mass stars.
Potentially, this could be alleviated by including a delay after sinks form before converting them to stars, so the sinks have more time to accrete gas and growm but we don't consider this further here.
In summary, setting the upper star mass limit equal to the group mass increases the sink-to-star conversion, but the resultant mass distribution is heavily skewed towards the lower mass end. 

\section{Conclusion}
\label{sec:conclusion}

Sink particles are used in star-forming simulations of various length scales and resolutions to replace dense gas regions, however a method to convert sinks to stars is needed if the sinks are not well resolved as individual stars, which is essential in simulations that require stellar properties like stellar feedback simulations. 
\cite{wall_flash_2019} introduced the \textit{single} star formation prescription, where a population of stars that is sampled from the input initial mass function (IMF) is introduced for each sink. 
This method works well with cluster sinks, but undersamples high mass stars if used on smaller mass sinks that are usually approximated as small groups of stars. 
In this paper, we introduce \textit{grouped} star formation, a modification of the \textit{single} star formation method whereby sinks are first grouped according to their positions, velocities, and ages, then the group masses are used to sample the IMF and form stars.
Using \texttt{Ekster} \citep{rieder_ekster_2021}, we test this method in simulations of various physical scales, from a sub-parsec isolated cloud simulation up to a kiloparsec spiral arm simulation.
We show that the \textit{grouped} star formation prescription is robust in simulations of different physical scales, and is essential in parsec-scale or smaller scale simulations, whereby their typical mass resolution of $10^{-4} - 10^{-1}$ \msun{} means that each sink is only approximated as a small star-forming region. 
This method would allow us to study the evolution of star clusters more accurately.

One of the main advantages of increasing the degree of  grouping of sinks is that the IMF is more complete, i.e. higher mass stars are sampled, as the degree of grouping increases. 
The disadvantage of increasing the grouping is that local mass may be less well conserved, e.g. stars are placed more preferentially at the few larger sinks
rather than spread out in the simulation. Furthermore stars formed at different times could be grouped together but this is easily constrained by our age criterion. For the models we present here, generally the `turbulent’ grouping scenario is optimal. For this we set the grouping distance $d\_g = 1$ pc, grouping speed $v\_g$ equal to the turbulence of the system, and grouping age $\tau\_g$ equal to the free-fall timescale. This tends to sample the IMF well, group sinks which form within a free-fall time, and reproduces most of the expected stellar mass in stars. This model is also more physically motivated, since the values of these parameters are approximately the physical scales expected in a typical star-forming region. 
. 

For our smallest scale region ($\sim 0.5$ pc), the `all grouping' method which groups all sinks as one produces the optimal results. This is not really surprising since this simulation is simply of one small star-forming region. We also find that we can produce a closer match to the velocity distribution of the stars formed in Bate 2012 by not including softening for the star particles, which again is not that surprising. With our grouping method we can basically reproduce the IMF, velocity characteristic of the stars, whilst the detailed spatial distribution of the stars is not that important since it is unlikely to be resolved in larger scale simulations. Our `turbulent' grouping model also works well, in this case this model is not very different to the `all grouping’ case - again we would expect the ‘turbulent’ grouping to group most of the sinks together if the `turbulent’ method is tuned to select individual star-forming regions (where stars located together form together on a free-fall time). The `no grouping’ model i.e. \textit{single} star formation prescription, is a particularly poor choice since only a small mass of the sinks are converted to stars, and there is an absence of massive stars. 

With low mass groups, which may arise in particularly small scale simulations, mass is not necessarily efficiently converted from sinks to stars, which is not satisfactory if we are treating the stars as the stellar component of the simulation.
We explored changing the upper limit of the IMF from a fixed value to the individual group masses, so that the fraction of mass in stars is increased, but the latter method also has the effect of suppressing the stellar masses such that low mass stars are overpopulated and high mass stars underpopulated. Using the ‘all grouping’ or `turbulent’ modes seem to be much better choices than the `no grouping' model here in any case.

In our intermediate scale simulations ($\sim 10$ pc), again the `all grouping' model samples the IMF well, but we note that on these scales it is not necessarily appropriate, or desirable to group all the sinks together. Again the `turbulent’ grouping still samples the IMF reasonably well, and better groups together sinks based on location,  and on these scales, age is also relevant. We note that potentially it may not actually be optimal to sample small clusters of stars according to a full IMF, as such regions may be less likely to contain massive stars, but we don’t consider this further here. 
Again the `no grouping’ case does not produce the IMF well, and has a lower fraction of mass in stars compared to the other cases.

Finally we checked our grouping method for a larger scale simulation ($\sim 1$ kpc), and verified that the `no grouping’ case, i.e. \textit{single} star formation is sufficient and there is no need to adopt grouping of sink particles once the mass of the sinks is sufficient to sample the IMF.

The type of IMF is a free parameter in our simulations.
We choose the Kroupa IMF \citep{kroupa_2001} for convenience, but one can also choose other types of IMF to suit the problem in hand, e.g the simple broken power law IMF by \cite{salpeter_imf_1955} or \cite{miller_initial_1979}, the IMF by \cite{chabrier_imf_2003} which models better the stars at lower mass end, and the IMF from the simulation by \cite{susa_mass_2014}, also investigated in \cite{hirai_sirius1_2021}, to form stars in the range of 1 - 300 \msun{} in cosmological simulations. 
We can only compare the resultant IMF of the stars with our selected choice of IMF.

Our algorithm is simple and easy to add in any star formation code, but it introduces three additional parameters to the simulation. 
We briefly explored other implementations like using the virial theorem to group the sinks, or reassigning group indices at every timestep, but found these algorithms are unable to group the sinks effectively.
Other recently-developed clustering methods \citep[e.g. ][]{torniamenti_clustering_2021} can potentially be adapted into our \textit{grouped} star formation prescription to be applied in hydrodynamical simulations.
One potential improvement which we didn't test further is allowing a delay, so that sink groups can accumulate mass before being converted to stars, although this has the disadvantage of introducing a further free parameter. Essentially though, our method can be easily refined to suit the needs of the user.

\section*{Acknowledgements}

The authors thank Tim Naylor for the helpful discussion on statistics, and Nicholas Herrington for the initial discussion on the analysis of B12 simulation comparison. SR acknowledges funding from STFC Consolidated Grant ST/R000395/1. CLD acknowledges funding from the European Research Council for the Horizon 2020 ERC consolidator grant project ICYBOB, grant number 818940. SEJ acknowledges support from the STFC grant ST/R000905/1 Simulations in this paper were performed on ISCA, University of Exeter. The column density plot of B12 cluster was produced using \texttt{SPLASH} \citep{price_splash_2007}, while those of cloud-cloud collision simulations were produced with the help of \texttt{Fi} \citep{pelupessy_fi_2004} in \texttt{AMUSE}.
The underlying gas velocity fields used to check the stellar velocity were computed using \texttt{Gadget2} \citep{volker_gadget2_2005} in \texttt{AMUSE}.
\texttt{Python} packages, such as \texttt{NumPy} \citep{python_numpy_2020}, \texttt{Matplotlib} \citep{python_matplotlib_2007}, \texttt{pandas} \citep{python_pandas1_2010}, and \texttt{SciPy} \citep{python_scipy_2020} were used generously to analyse the results in this paper.  

\section*{Data Availability}

The data underlying this paper will be shared on reasonable request to the corresponding author.

\bibliographystyle{mnras}
\bibliography{mybib}

\begin{thebibliography}{}
\makeatletter
\relax
\def\mn@urlcharsother{\let\do\@makeother \do\$\do\&\do\#\do\^\do\_\do\%\do\~}
\def\mn@doi{\begingroup\mn@urlcharsother \@ifnextchar [ {\mn@doi@}
  {\mn@doi@[]}}
\def\mn@doi@[#1]#2{\def\@tempa{#1}\ifx\@tempa\@empty \href
  {http://dx.doi.org/#2} {doi:#2}\else \href {http://dx.doi.org/#2} {#1}\fi
  \endgroup}
\def\mn@eprint#1#2{\mn@eprint@#1:#2::\@nil}
\def\mn@eprint@arXiv#1{\href {http://arxiv.org/abs/#1} {{\tt arXiv:#1}}}
\def\mn@eprint@dblp#1{\href {http://dblp.uni-trier.de/rec/bibtex/#1.xml}
  {dblp:#1}}
\def\mn@eprint@#1:#2:#3:#4\@nil{\def\@tempa {#1}\def\@tempb {#2}\def\@tempc
  {#3}\ifx \@tempc \@empty \let \@tempc \@tempb \let \@tempb \@tempa \fi \ifx
  \@tempb \@empty \def\@tempb {arXiv}\fi \@ifundefined
  {mn@eprint@\@tempb}{\@tempb:\@tempc}{\expandafter \expandafter \csname
  mn@eprint@\@tempb\endcsname \expandafter{\@tempc}}}

\bibitem[\protect\citeauthoryear{{Ali}}{{Ali}}{2021}]{ali_hiiregion_2021}
{Ali} A.~A.,  2021, \mn@doi [\mnras] {10.1093/mnras/staa3992}, \href
  {https://ui.adsabs.harvard.edu/abs/2021MNRAS.501.4136A} {501, 4136}

\bibitem[\protect\citeauthoryear{{Ali} \& {Harries}}{{Ali} \&
  {Harries}}{2019}]{ali_feedback_2019}
{Ali} A.~A.,  {Harries} T.~J.,  2019, \mn@doi [\mnras] {10.1093/mnras/stz1673},
  \href {https://ui.adsabs.harvard.edu/abs/2019MNRAS.487.4890A} {487, 4890}

\bibitem[\protect\citeauthoryear{Balfour, Whitworth, Hubber  \& Jaffa}{Balfour
  et~al.}{2015}]{balfour_star_2015}
Balfour S.~K.,  Whitworth A.~P.,  Hubber D.~A.,   Jaffa S.~E.,  2015, \mn@doi
  [MNRAS] {10.1093/mnras/stv1772}, 453, 2472

\bibitem[\protect\citeauthoryear{{Ballone}, {Torniamenti}, {Mapelli}, {Di
  Carlo}, {Spera}, {Rastello}, {Gaspari}  \& {Iorio}}{{Ballone}
  et~al.}{2021}]{ballone_stars_2021}
{Ballone} A.,  {Torniamenti} S.,  {Mapelli} M.,  {Di Carlo} U.~N.,  {Spera} M.,
   {Rastello} S.,  {Gaspari} N.,   {Iorio} G.,  2021, \mn@doi [\mnras]
  {10.1093/mnras/staa3763}, \href
  {https://ui.adsabs.harvard.edu/abs/2021MNRAS.501.2920B} {501, 2920}

\bibitem[\protect\citeauthoryear{{Bate}}{{Bate}}{2009}]{bate_cluster_2009}
{Bate} M.~R.,  2009, \mn@doi [\mnras] {10.1111/j.1365-2966.2008.14106.x}, \href
  {https://ui.adsabs.harvard.edu/abs/2009MNRAS.392..590B} {392, 590}

\bibitem[\protect\citeauthoryear{{Bate}}{{Bate}}{2012}]{bate_cluster_2012}
{Bate} M.~R.,  2012, \mn@doi [\mnras] {10.1111/j.1365-2966.2011.19955.x}, \href
  {https://ui.adsabs.harvard.edu/abs/2012MNRAS.419.3115B} {419, 3115}

\bibitem[\protect\citeauthoryear{Bate \& Burkert}{Bate \&
  Burkert}{1997}]{bate_resolution_1997}
Bate M.~R.,  Burkert A.,  1997, \mn@doi [\mnras] {10.1093/mnras/288.4.1060},
  288, 1060

\bibitem[\protect\citeauthoryear{Bate, Bonnell  \& Price}{Bate
  et~al.}{1995}]{bate_modelling_1995}
Bate M.~R.,  Bonnell I.~A.,   Price N.~M.,  1995, \mn@doi [MNRAS]
  {10.1093/mnras/277.2.362}, 277, 362

\bibitem[\protect\citeauthoryear{Bate, Bonnell  \& Bromm}{Bate
  et~al.}{2003}]{bate_formation_2003}
Bate M.~R.,  Bonnell I.~A.,   Bromm V.,  2003, \mn@doi [MNRAS]
  {10.1046/j.1365-8711.2003.06210.x}, 339, 577

\bibitem[\protect\citeauthoryear{{Bending}, {Dobbs}  \& {Bate}}{{Bending}
  et~al.}{2020}]{bending_photoionisation_2020}
{Bending} T. J.~R.,  {Dobbs} C.~L.,   {Bate} M.~R.,  2020, \mn@doi [\mnras]
  {10.1093/mnras/staa1293}, \href
  {https://ui.adsabs.harvard.edu/abs/2020MNRAS.495.1672B} {495, 1672}

\bibitem[\protect\citeauthoryear{{Bertelli Motta}, {Clark}, {Glover}, {Klessen}
   \& {Pasquali}}{{Bertelli Motta} et~al.}{2016}]{bertellimotta_imf_2016}
{Bertelli Motta} C.,  {Clark} P.~C.,  {Glover} S.~C.~O.,  {Klessen} R.~S.,
  {Pasquali} A.,  2016, \mn@doi [\mnras] {10.1093/mnras/stw1921}, \href
  {https://ui.adsabs.harvard.edu/abs/2016MNRAS.462.4171B} {462, 4171}

\bibitem[\protect\citeauthoryear{Bleuler \& Teyssier}{Bleuler \&
  Teyssier}{2014}]{bleuler_towards_2014}
Bleuler A.,  Teyssier R.,  2014, \mn@doi [Monthly Notices of the Royal
  Astronomical Society] {10.1093/mnras/stu2005}, 445, 4015

\bibitem[\protect\citeauthoryear{{Bonnell}, {Bate}, {Clarke}  \&
  {Pringle}}{{Bonnell} et~al.}{1997}]{bonnell_accretion_1997}
{Bonnell} I.~A.,  {Bate} M.~R.,  {Clarke} C.~J.,   {Pringle} J.~E.,  1997,
  \mn@doi [\mnras] {10.1093/mnras/285.1.201}, \href
  {https://ui.adsabs.harvard.edu/abs/1997MNRAS.285..201B} {285, 201}

\bibitem[\protect\citeauthoryear{{Boss}}{{Boss}}{2019}]{boss_protostar_2019}
{Boss} A.~P.,  2019, \mn@doi [\apj] {10.3847/1538-4357/aaf005}, \href
  {https://ui.adsabs.harvard.edu/abs/2019ApJ...870....3B} {870, 3}

\bibitem[\protect\citeauthoryear{{Chabrier}}{{Chabrier}}{2003}]{chabrier_imf_2003}
{Chabrier} G.,  2003, \mn@doi [\pasp] {10.1086/376392}, \href
  {https://ui.adsabs.harvard.edu/abs/2003PASP..115..763C} {115, 763}

\bibitem[\protect\citeauthoryear{{Dale}, {W{\"u}nsch}, {Smith}, {Whitworth}  \&
  {Palou{\v{s}}}}{{Dale} et~al.}{2011}]{dale_accretion_2011}
{Dale} J.~E.,  {W{\"u}nsch} R.,  {Smith} R.~J.,  {Whitworth} A.,
  {Palou{\v{s}}} J.,  2011, \mn@doi [\mnras]
  {10.1111/j.1365-2966.2010.17844.x}, \href
  {https://ui.adsabs.harvard.edu/abs/2011MNRAS.411.2230D} {411, 2230}

\bibitem[\protect\citeauthoryear{{Dale}, {Ngoumou}, {Ercolano}  \&
  {Bonnell}}{{Dale} et~al.}{2014}]{dale_feedback_2014}
{Dale} J.~E.,  {Ngoumou} J.,  {Ercolano} B.,   {Bonnell} I.~A.,  2014, \mn@doi
  [\mnras] {10.1093/mnras/stu816}, \href
  {https://ui.adsabs.harvard.edu/abs/2014MNRAS.442..694D} {442, 694}

\bibitem[\protect\citeauthoryear{Dinnbier \& Walch}{Dinnbier \&
  Walch}{2020}]{dinnbier_feedback_2020}
Dinnbier F.,  Walch S.,  2020, \mn@doi [Monthly Notices of the Royal
  Astronomical Society] {10.1093/mnras/staa2560}, 499, 748

\bibitem[\protect\citeauthoryear{{Dobbs} \& {Wurster}}{{Dobbs} \&
  {Wurster}}{2021}]{dobbs_magnetic_YMC_2021}
{Dobbs} C.~L.,  {Wurster} J.,  2021, \mn@doi [\mnras] {10.1093/mnras/stab150},
  \href {https://ui.adsabs.harvard.edu/abs/2021MNRAS.502.2285D} {502, 2285}

\bibitem[\protect\citeauthoryear{{Dobbs}, {Liow}  \& {Rieder}}{{Dobbs}
  et~al.}{2020}]{dobbs_ymc_2020}
{Dobbs} C.~L.,  {Liow} K.~Y.,   {Rieder} S.,  2020, \mn@doi [\mnras]
  {10.1093/mnrasl/slaa072}, \href
  {https://ui.adsabs.harvard.edu/abs/2020MNRAS.496L...1D} {496, L1}

\bibitem[\protect\citeauthoryear{Ester, Kriegel, Sander  \& Xu}{Ester
  et~al.}{1996}]{ester_density_based_1996}
Ester M.,  Kriegel H.-P.,  Sander J.,   Xu X.,  1996, in Proceedings of the
  Second International Conference on Knowledge Discovery and Data Mining.
  KDD’96.
AAAI Press, p. 226–231

\bibitem[\protect\citeauthoryear{{Federrath} \& {Klessen}}{{Federrath} \&
  {Klessen}}{2012}]{federrath_sfr_2012}
{Federrath} C.,  {Klessen} R.~S.,  2012, \mn@doi [\apj]
  {10.1088/0004-637X/761/2/156}, \href
  {https://ui.adsabs.harvard.edu/abs/2012ApJ...761..156F} {761, 156}

\bibitem[\protect\citeauthoryear{Federrath, Banerjee, Clark  \&
  Klessen}{Federrath et~al.}{2010}]{federrath_modeling_2010}
Federrath C.,  Banerjee R.,  Clark P.~C.,   Klessen R.~S.,  2010, \mn@doi [The
  Astrophysical Journal] {10.1088/0004-637X/713/1/269}, 713, 269

\bibitem[\protect\citeauthoryear{Fujii \& Portegies~Zwart}{Fujii \&
  Portegies~Zwart}{2015}]{fujii_initial_2015}
Fujii M.~S.,  Portegies~Zwart S.,  2015, \mn@doi [\mnras]
  {10.1093/mnras/stv293}, 449, 726

\bibitem[\protect\citeauthoryear{{Fujii}, {Iwasawa}, {Funato}  \&
  {Makino}}{{Fujii} et~al.}{2007}]{fujii_bridge_2007}
{Fujii} M.,  {Iwasawa} M.,  {Funato} Y.,   {Makino} J.,  2007, \mn@doi [\pasj]
  {10.1093/pasj/59.6.1095}, \href
  {https://ui.adsabs.harvard.edu/abs/2007PASJ...59.1095F} {59, 1095}

\bibitem[\protect\citeauthoryear{{Geen}, {Hennebelle}, {Tremblin}  \&
  {Rosdahl}}{{Geen} et~al.}{2016}]{geen_feedback_2016}
{Geen} S.,  {Hennebelle} P.,  {Tremblin} P.,   {Rosdahl} J.,  2016, \mn@doi
  [\mnras] {10.1093/mnras/stw2235}, \href
  {https://ui.adsabs.harvard.edu/abs/2016MNRAS.463.3129G} {463, 3129}

\bibitem[\protect\citeauthoryear{{Geen}, {Watson}, {Rosdahl}, {Bieri},
  {Klessen}  \& {Hennebelle}}{{Geen} et~al.}{2018}]{geen_star_formation_2018}
{Geen} S.,  {Watson} S.~K.,  {Rosdahl} J.,  {Bieri} R.,  {Klessen} R.~S.,
  {Hennebelle} P.,  2018, \mn@doi [\mnras] {10.1093/mnras/sty2439}, \href
  {https://ui.adsabs.harvard.edu/abs/2018MNRAS.481.2548G} {481, 2548}

\bibitem[\protect\citeauthoryear{{Girichidis}, {Federrath}, {Banerjee}  \&
  {Klessen}}{{Girichidis} et~al.}{2011}]{girichidis_initial_2011}
{Girichidis} P.,  {Federrath} C.,  {Banerjee} R.,   {Klessen} R.~S.,  2011,
  \mn@doi [\mnras] {10.1111/j.1365-2966.2011.18348.x}, \href
  {https://ui.adsabs.harvard.edu/abs/2011MNRAS.413.2741G} {413, 2741}

\bibitem[\protect\citeauthoryear{G{\'o}mez-de Mariscal, Guerrero, Sneider,
  Jayatilaka, Phillip, Wirtz  \& Mu{\~n}oz-Barrutia}{G{\'o}mez-de Mariscal
  et~al.}{2021}]{phacking_large_sample_2021}
G{\'o}mez-de Mariscal E.,  Guerrero V.,  Sneider A.,  Jayatilaka H.,  Phillip
  J.~M.,  Wirtz D.,   Mu{\~n}oz-Barrutia A.,  2021, \mn@doi [bioRxiv]
  {10.1101/2019.12.17.878405}

\bibitem[\protect\citeauthoryear{{Greif}, {Bromm}, {Clark}, {Glover}, {Smith},
  {Klessen}, {Yoshida}  \& {Springel}}{{Greif}
  et~al.}{2012}]{greif_protostellar_2012}
{Greif} T.~H.,  {Bromm} V.,  {Clark} P.~C.,  {Glover} S. C.~O.,  {Smith} R.~J.,
   {Klessen} R.~S.,  {Yoshida} N.,   {Springel} V.,  2012, \mn@doi [\mnras]
  {10.1111/j.1365-2966.2012.21212.x}, \href
  {https://ui.adsabs.harvard.edu/abs/2012MNRAS.424..399G} {424, 399}

\bibitem[\protect\citeauthoryear{{Grudi{\'c}}, {Guszejnov}, {Hopkins}, {Offner}
   \& {Faucher-Gigu{\`e}re}}{{Grudi{\'c}} et~al.}{2021}]{grudic_starforge_2021}
{Grudi{\'c}} M.~Y.,  {Guszejnov} D.,  {Hopkins} P.~F.,  {Offner} S. S.~R.,
  {Faucher-Gigu{\`e}re} C.-A.,  2021, \mn@doi [\mnras]
  {10.1093/mnras/stab1347}, \href
  {https://ui.adsabs.harvard.edu/abs/2021MNRAS.506.2199G} {506, 2199}

\bibitem[\protect\citeauthoryear{Harris et~al.,}{Harris
  et~al.}{2020}]{python_numpy_2020}
Harris C.~R.,  et~al., 2020, \mn@doi [Nature] {10.1038/s41586-020-2649-2}, 585,
  357

\bibitem[\protect\citeauthoryear{{He}, {Ricotti}  \& {Geen}}{{He}
  et~al.}{2019}]{he_clusters_2019}
{He} C.-C.,  {Ricotti} M.,   {Geen} S.,  2019, \mn@doi [\mnras]
  {10.1093/mnras/stz2239}, \href
  {https://ui.adsabs.harvard.edu/abs/2019MNRAS.489.1880H} {489, 1880}

\bibitem[\protect\citeauthoryear{{Hennebelle}, {Commer{\c{c}}on}, {Lee}  \&
  {Charnoz}}{{Hennebelle} et~al.}{2020}]{hennebelle_disc_2020}
{Hennebelle} P.,  {Commer{\c{c}}on} B.,  {Lee} Y.-N.,   {Charnoz} S.,  2020,
  \mn@doi [\aap] {10.1051/0004-6361/201936714}, \href
  {https://ui.adsabs.harvard.edu/abs/2020A&A...635A..67H} {635, A67}

\bibitem[\protect\citeauthoryear{Hirai, Fujii  \& Saitoh}{Hirai
  et~al.}{2021}]{hirai_sirius1_2021}
Hirai Y.,  Fujii M.~S.,   Saitoh T.~R.,  2021, arXiv:2005.12906 [astro-ph]

\bibitem[\protect\citeauthoryear{{Hislop}, {Naab}, {Steinwandel}, {Lah{\'e}n},
  {Irodotou}, {Johansson}  \& {Walch}}{{Hislop}
  et~al.}{2021}]{hislop_dwarf_2021}
{Hislop} J.~M.,  {Naab} T.,  {Steinwandel} U.~P.,  {Lah{\'e}n} N.,  {Irodotou}
  D.,  {Johansson} P.~H.,   {Walch} S.,  2021, arXiv e-prints, \href
  {https://ui.adsabs.harvard.edu/abs/2021arXiv210908160H} {p. arXiv:2109.08160}

\bibitem[\protect\citeauthoryear{{Howard}, {Pudritz}  \& {Harris}}{{Howard}
  et~al.}{2018}]{howard_YMC_2018}
{Howard} C.~S.,  {Pudritz} R.~E.,   {Harris} W.~E.,  2018, \mn@doi [Nature
  Astronomy] {10.1038/s41550-018-0506-0}, \href
  {https://ui.adsabs.harvard.edu/abs/2018NatAs...2..725H} {2, 725}

\bibitem[\protect\citeauthoryear{{Hu}, {Naab}, {Glover}, {Walch}  \&
  {Clark}}{{Hu} et~al.}{2017}]{hu_interstellar_2017}
{Hu} C.-Y.,  {Naab} T.,  {Glover} S. C.~O.,  {Walch} S.,   {Clark} P.~C.,
  2017, \mn@doi [\mnras] {10.1093/mnras/stx1773}, \href
  {https://ui.adsabs.harvard.edu/abs/2017MNRAS.471.2151H} {471, 2151}

\bibitem[\protect\citeauthoryear{{Hubber}, {Walch}  \& {Whitworth}}{{Hubber}
  et~al.}{2013}]{hubber_sink_2013}
{Hubber} D.~A.,  {Walch} S.,   {Whitworth} A.~P.,  2013, \mn@doi [MNRAS]
  {10.1093/mnras/stt128}, \href
  {https://ui.adsabs.harvard.edu/abs/2013MNRAS.430.3261H} {430, 3261}

\bibitem[\protect\citeauthoryear{{Hubber}, {Rosotti}  \& {Booth}}{{Hubber}
  et~al.}{2018}]{hubber_gandalf_2018}
{Hubber} D.~A.,  {Rosotti} G.~P.,   {Booth} R.~A.,  2018, \mn@doi [\mnras]
  {10.1093/mnras/stx2405}, \href
  {https://ui.adsabs.harvard.edu/abs/2018MNRAS.473.1603H} {473, 1603}

\bibitem[\protect\citeauthoryear{Hunter}{Hunter}{2007}]{python_matplotlib_2007}
Hunter J.~D.,  2007, \mn@doi [Computing in Science \& Engineering]
  {10.1109/MCSE.2007.55}, 9, 90

\bibitem[\protect\citeauthoryear{{Jaffa}}{{Jaffa}}{prep}]{jaffa_sims_prep}
{Jaffa} S.,  in prep., some journal

\bibitem[\protect\citeauthoryear{{Jones} \& {Bate}}{{Jones} \&
  {Bate}}{2018}]{jones_radiative_2018}
{Jones} M.~O.,  {Bate} M.~R.,  2018, \mn@doi [\mnras] {10.1093/mnras/sty1969},
  \href {https://ui.adsabs.harvard.edu/abs/2018MNRAS.480.2562J} {480, 2562}

\bibitem[\protect\citeauthoryear{{Kroupa}}{{Kroupa}}{2001}]{kroupa_2001}
{Kroupa} P.,  2001, \mn@doi [\mnras] {10.1046/j.1365-8711.2001.04022.x}, \href
  {https://ui.adsabs.harvard.edu/abs/2001MNRAS.322..231K} {322, 231}

\bibitem[\protect\citeauthoryear{{Kuznetsova}, {Hartmann}  \&
  {Heitsch}}{{Kuznetsova} et~al.}{2020}]{kuznetsova_protostellar_2020}
{Kuznetsova} A.,  {Hartmann} L.,   {Heitsch} F.,  2020, \mn@doi [\apj]
  {10.3847/1538-4357/ab7eac}, \href
  {https://ui.adsabs.harvard.edu/abs/2020ApJ...893...73K} {893, 73}

\bibitem[\protect\citeauthoryear{{Lah{\'e}n}, {Naab}, {Johansson}, {Elmegreen},
  {Hu}, {Walch}, {Steinwandel}  \& {Moster}}{{Lah{\'e}n}
  et~al.}{2020}]{lahen_griffin_2020}
{Lah{\'e}n} N.,  {Naab} T.,  {Johansson} P.~H.,  {Elmegreen} B.,  {Hu} C.-Y.,
  {Walch} S.,  {Steinwandel} U.~P.,   {Moster} B.~P.,  2020, \mn@doi [\apj]
  {10.3847/1538-4357/ab7190}, \href
  {https://ui.adsabs.harvard.edu/abs/2020ApJ...891....2L} {891, 2}

\bibitem[\protect\citeauthoryear{{Latif} \& {Volonteri}}{{Latif} \&
  {Volonteri}}{2015}]{latif_black_holes_2015}
{Latif} M.~A.,  {Volonteri} M.,  2015, \mn@doi [\mnras]
  {10.1093/mnras/stv1337}, \href
  {https://ui.adsabs.harvard.edu/abs/2015MNRAS.452.1026L} {452, 1026}

\bibitem[\protect\citeauthoryear{{Liow} \& {Dobbs}}{{Liow} \&
  {Dobbs}}{2020}]{liow_collision_2020}
{Liow} K.~Y.,  {Dobbs} C.~L.,  2020, \mn@doi [\mnras] {10.1093/mnras/staa2857},
  \href {https://ui.adsabs.harvard.edu/abs/2020MNRAS.499.1099L} {499, 1099}

\bibitem[\protect\citeauthoryear{{Lomax}, {Whitworth}  \& {Hubber}}{{Lomax}
  et~al.}{2015}]{lomax_prestellar_2015}
{Lomax} O.,  {Whitworth} A.~P.,   {Hubber} D.~A.,  2015, \mn@doi [\mnras]
  {10.1093/mnras/stv310}, \href
  {https://ui.adsabs.harvard.edu/abs/2015MNRAS.449..662L} {449, 662}

\bibitem[\protect\citeauthoryear{Masunaga \& Inutsuka}{Masunaga \&
  Inutsuka}{2000}]{Masunaga_2000}
Masunaga H.,  Inutsuka S.-i.,  2000, \mn@doi [ApJ] {10.1086/308901}, 536, 406

\bibitem[\protect\citeauthoryear{Masunaga, Miyama  \& Inutsuka}{Masunaga
  et~al.}{1998}]{masunaga_collapse_1998}
Masunaga H.,  Miyama S.~M.,   Inutsuka S.-i.,  1998, \mn@doi [The Astrophysical
  Journal] {10.1086/305281}, 495, 346

\bibitem[\protect\citeauthoryear{Miller \& Scalo}{Miller \&
  Scalo}{1979}]{miller_initial_1979}
Miller G.~E.,  Scalo J.~M.,  1979, \mn@doi [ApJS] {10.1086/190629}, 41, 513

\bibitem[\protect\citeauthoryear{{Monaghan}}{{Monaghan}}{1997}]{monaghan_1997}
{Monaghan} J.~J.,  1997, \mn@doi [J. Comput. Phys.] {10.1006/jcph.1997.5732},
  \href {http://adsabs.harvard.edu/abs/1997JCoPh.136..298M} {136, 298}

\bibitem[\protect\citeauthoryear{{Morris} \& {Monaghan}}{{Morris} \&
  {Monaghan}}{1997}]{morris_1997}
{Morris} J.~P.,  {Monaghan} J.~J.,  1997, \mn@doi [J. Comput. Phys.]
  {10.1006/jcph.1997.5690}, \href
  {http://adsabs.harvard.edu/abs/1997JCoPh.136...41M} {136, 41}

\bibitem[\protect\citeauthoryear{{Ntormousi} \& {Hennebelle}}{{Ntormousi} \&
  {Hennebelle}}{2019}]{ntormousi_filaments_2019}
{Ntormousi} E.,  {Hennebelle} P.,  2019, \mn@doi [\aap]
  {10.1051/0004-6361/201834094}, \href
  {https://ui.adsabs.harvard.edu/abs/2019A&A...625A..82N} {625, A82}

\bibitem[\protect\citeauthoryear{{Padoan} \& {Nordlund}}{{Padoan} \&
  {Nordlund}}{2011}]{padoan_turbulence_2011}
{Padoan} P.,  {Nordlund} {\r{A}}.,  2011, \mn@doi [\apj]
  {10.1088/0004-637X/730/1/40}, \href
  {https://ui.adsabs.harvard.edu/abs/2011ApJ...730...40P} {730, 40}

\bibitem[\protect\citeauthoryear{{Pelupessy}, {van der Werf}  \&
  {Icke}}{{Pelupessy} et~al.}{2004}]{pelupessy_fi_2004}
{Pelupessy} F.~I.,  {van der Werf} P.~P.,   {Icke} V.,  2004, \mn@doi [\aap]
  {10.1051/0004-6361:20047071}, \href
  {https://ui.adsabs.harvard.edu/abs/2004A&A...422...55P} {422, 55}

\bibitem[\protect\citeauthoryear{{Perets} \& {{\v{S}}ubr}}{{Perets} \&
  {{\v{S}}ubr}}{2012}]{perets_runaways_2012}
{Perets} H.~B.,  {{\v{S}}ubr} L.,  2012, \mn@doi [\apj]
  {10.1088/0004-637X/751/2/133}, \href
  {https://ui.adsabs.harvard.edu/abs/2012ApJ...751..133P} {751, 133}

\bibitem[\protect\citeauthoryear{{Portegies Zwart} \& {Verbunt}}{{Portegies
  Zwart} \& {Verbunt}}{1996}]{portegies_zwart_seba1_1996}
{Portegies Zwart} S.~F.,  {Verbunt} F.,  1996, \aap, \href
  {https://ui.adsabs.harvard.edu/abs/1996A&A...309..179P} {309, 179}

\bibitem[\protect\citeauthoryear{{Portegies Zwart} et~al.,}{{Portegies Zwart}
  et~al.}{2018}]{amuse_2018}
{Portegies Zwart} S.,  et~al., 2018, {Amuse: The Astrophysical Multipurpose
  Software Environment}, \mn@doi{10.5281/zenodo.1443252}

\bibitem[\protect\citeauthoryear{Price}{Price}{2007}]{price_splash_2007}
Price D.~J.,  2007, \mn@doi [Publ. Astron. Soc. Australia] {10.1071/AS07022},
  24, 159–173

\bibitem[\protect\citeauthoryear{{Price} \& {Bate}}{{Price} \&
  {Bate}}{2008}]{price_magnetic_2008}
{Price} D.~J.,  {Bate} M.~R.,  2008, \mn@doi [\mnras]
  {10.1111/j.1365-2966.2008.12976.x}, \href
  {https://ui.adsabs.harvard.edu/abs/2008MNRAS.385.1820P} {385, 1820}

\bibitem[\protect\citeauthoryear{{Price} \& {Federrath}}{{Price} \&
  {Federrath}}{2010}]{price_comparison_2010}
{Price} D.~J.,  {Federrath} C.,  2010, \mn@doi [\mnras]
  {10.1111/j.1365-2966.2010.16810.x}, \href
  {https://ui.adsabs.harvard.edu/abs/2010MNRAS.406.1659P} {406, 1659}

\bibitem[\protect\citeauthoryear{{Price} \& {Monaghan}}{{Price} \&
  {Monaghan}}{2007}]{price_gravity_2007}
{Price} D.~J.,  {Monaghan} J.~J.,  2007, \mn@doi [MNRAS]
  {10.1111/j.1365-2966.2006.11241.x}, \href
  {http://adsabs.harvard.edu/abs/2007MNRAS.374.1347P} {374, 1347}

\bibitem[\protect\citeauthoryear{{Price} et~al.,}{{Price}
  et~al.}{2018}]{price_phantom_2018}
{Price} D.~J.,  et~al., 2018, \mn@doi [PASA] {10.1017/pasa.2018.25}, \href
  {https://ui.adsabs.harvard.edu/abs/2018PASA...35...31P} {35, 31}

\bibitem[\protect\citeauthoryear{Præstgaard}{Præstgaard}{1995}]{kstest_permutation_1995}
Præstgaard J.~T.,  1995, Scandinavian Journal of Statistics, 22, 305

\bibitem[\protect\citeauthoryear{{Renaud}, {Bournaud}  \& {Duc}}{{Renaud}
  et~al.}{2015}]{renaud_antennae_2015}
{Renaud} F.,  {Bournaud} F.,   {Duc} P.-A.,  2015, \mn@doi [\mnras]
  {10.1093/mnras/stu2208}, \href
  {https://ui.adsabs.harvard.edu/abs/2015MNRAS.446.2038R} {446, 2038}

\bibitem[\protect\citeauthoryear{Rieder \& Liow}{Rieder \&
  Liow}{2021}]{ekster_code}
Rieder S.,  Liow K.~Y.,  2021, rieder/ekster:, \mn@doi{10.5281/zenodo.5520944},
  \url {https://doi.org/10.5281/zenodo.5520944}

\bibitem[\protect\citeauthoryear{{Rieder}, {Dobbs}, {Bending}, {Liow}  \&
  {Wurster}}{{Rieder} et~al.}{2021}]{rieder_ekster_2021}
{Rieder} S.,  {Dobbs} C.,  {Bending} T.,  {Liow} K.~Y.,   {Wurster} J.,  2021,
  arXiv e-prints, \href {https://ui.adsabs.harvard.edu/abs/2021arXiv211109720R}
  {p. arXiv:2111.09720}

\bibitem[\protect\citeauthoryear{{Saitoh} \& {Makino}}{{Saitoh} \&
  {Makino}}{2013}]{saitoh_disph_2013}
{Saitoh} T.~R.,  {Makino} J.,  2013, \mn@doi [\apj]
  {10.1088/0004-637X/768/1/44}, \href
  {https://ui.adsabs.harvard.edu/abs/2013ApJ...768...44S} {768, 44}

\bibitem[\protect\citeauthoryear{{Salpeter}}{{Salpeter}}{1955}]{salpeter_imf_1955}
{Salpeter} E.~E.,  1955, \mn@doi [\apj] {10.1086/145971}, \href
  {https://ui.adsabs.harvard.edu/abs/1955ApJ...121..161S} {121, 161}

\bibitem[\protect\citeauthoryear{Smith}{Smith}{2021}]{smith_sensitivity_2021}
Smith M.~C.,  2021, \mn@doi [Monthly Notices of the Royal Astronomical Society]
  {10.1093/mnras/stab291}, 502, 5417

\bibitem[\protect\citeauthoryear{{Sormani}, {Tre{\ss}}, {Klessen}  \&
  {Glover}}{{Sormani} et~al.}{2017}]{sormani_sinks_stars_2017}
{Sormani} M.~C.,  {Tre{\ss}} R.~G.,  {Klessen} R.~S.,   {Glover} S. C.~O.,
  2017, \mn@doi [\mnras] {10.1093/mnras/stw3205}, \href
  {https://ui.adsabs.harvard.edu/abs/2017MNRAS.466..407S} {466, 407}

\bibitem[\protect\citeauthoryear{{Springel}}{{Springel}}{2005}]{volker_gadget2_2005}
{Springel} V.,  2005, \mn@doi [\mnras] {10.1111/j.1365-2966.2005.09655.x},
  \href {https://ui.adsabs.harvard.edu/abs/2005MNRAS.364.1105S} {364, 1105}

\bibitem[\protect\citeauthoryear{{Stacy}, {Greif}  \& {Bromm}}{{Stacy}
  et~al.}{2010}]{stacy_multiples_2010}
{Stacy} A.,  {Greif} T.~H.,   {Bromm} V.,  2010, \mn@doi [\mnras]
  {10.1111/j.1365-2966.2009.16113.x}, \href
  {https://ui.adsabs.harvard.edu/abs/2010MNRAS.403...45S} {403, 45}

\bibitem[\protect\citeauthoryear{{Stamatellos}, {Whitworth}  \&
  {Hubber}}{{Stamatellos} et~al.}{2012}]{stamatellos_accretion_2012}
{Stamatellos} D.,  {Whitworth} A.~P.,   {Hubber} D.~A.,  2012, \mn@doi [\mnras]
  {10.1111/j.1365-2966.2012.22038.x}, \href
  {https://ui.adsabs.harvard.edu/abs/2012MNRAS.427.1182S} {427, 1182}

\bibitem[\protect\citeauthoryear{{Susa}, {Hasegawa}  \& {Tominaga}}{{Susa}
  et~al.}{2014}]{susa_mass_2014}
{Susa} H.,  {Hasegawa} K.,   {Tominaga} N.,  2014, \mn@doi [\apj]
  {10.1088/0004-637X/792/1/32}, \href
  {https://ui.adsabs.harvard.edu/abs/2014ApJ...792...32S} {792, 32}

\bibitem[\protect\citeauthoryear{{Torniamenti}, {Pasquato}, {Di Cintio},
  {Ballone}, {Iorio}  \& {Mapelli}}{{Torniamenti}
  et~al.}{2021}]{torniamenti_clustering_2021}
{Torniamenti} S.,  {Pasquato} M.,  {Di Cintio} P.,  {Ballone} A.,  {Iorio} G.,
   {Mapelli} M.,  2021, arXiv e-prints, \href
  {https://ui.adsabs.harvard.edu/abs/2021arXiv210600684T} {p. arXiv:2106.00684}

\bibitem[\protect\citeauthoryear{V{\'a}zquez{-}Semadeni, G{\'o}mez, Jappsen,
  Ballesteros{-}Paredes, Gonz{\'a}lez  \& Klessen}{V{\'a}zquez{-}Semadeni
  et~al.}{2007}]{vazquezsemadeni_molecular_2007}
V{\'a}zquez{-}Semadeni E.,  G{\'o}mez G.~C.,  Jappsen A.~K.,
  Ballesteros{-}Paredes J.,  Gonz{\'a}lez R.~F.,   Klessen R.~S.,  2007,
  \mn@doi [ApJ] {10.1086/510771}, 657, 870

\bibitem[\protect\citeauthoryear{Virtanen et~al.,}{Virtanen
  et~al.}{2020}]{python_scipy_2020}
Virtanen P.,  et~al., 2020, \mn@doi [Nature Methods]
  {10.1038/s41592-019-0686-2}, \href {https://rdcu.be/b08Wh} {17, 261}

\bibitem[\protect\citeauthoryear{{Wall}, {McMillan}, {Mac Low}, {Klessen}  \&
  {Portegies Zwart}}{{Wall} et~al.}{2019}]{wall_flash_2019}
{Wall} J.~E.,  {McMillan} S. L.~W.,  {Mac Low} M.-M.,  {Klessen} R.~S.,
  {Portegies Zwart} S.,  2019, \mn@doi [\apj] {10.3847/1538-4357/ab4db1}, \href
  {https://ui.adsabs.harvard.edu/abs/2019ApJ...887...62W} {887, 62}

\bibitem[\protect\citeauthoryear{{Wang}, {Iwasawa}, {Nitadori}  \&
  {Makino}}{{Wang} et~al.}{2020}]{wang_petar_2020}
{Wang} L.,  {Iwasawa} M.,  {Nitadori} K.,   {Makino} J.,  2020, \mn@doi
  [\mnras] {10.1093/mnras/staa1915}, \href
  {https://ui.adsabs.harvard.edu/abs/2020MNRAS.497..536W} {497, 536}

\bibitem[\protect\citeauthoryear{{W}es {M}c{K}inney}{{W}es
  {M}c{K}inney}{2010}]{python_pandas1_2010}
{W}es {M}c{K}inney 2010, in {S}t\'efan van~der {W}alt {J}arrod {M}illman eds,
  {P}roceedings of the 9th {P}ython in {S}cience {C}onference. pp 56 -- 61,
  \mn@doi{10.25080/Majora-92bf1922-00a}

\bibitem[\protect\citeauthoryear{{Whitehouse}, {Bate}  \&
  {Monaghan}}{{Whitehouse} et~al.}{2005}]{whitehouse_sph_radiation_2005}
{Whitehouse} S.~C.,  {Bate} M.~R.,   {Monaghan} J.~J.,  2005, \mn@doi [\mnras]
  {10.1111/j.1365-2966.2005.09683.x}, \href
  {https://ui.adsabs.harvard.edu/abs/2005MNRAS.364.1367W} {364, 1367}

\makeatother
\end{thebibliography}

\label{lastpage}
\end{document}